\documentclass[12pt]{article}

\usepackage{graphicx}
\usepackage{subcaption}
\usepackage{epsfig}
\usepackage{epstopdf}
\usepackage{setspace,caption}
\usepackage{xcolor}
\usepackage[T1]{fontenc}

\DeclareSymbolFont{myletters}{OML}{ztmcm}{m}{it}
\DeclareMathSymbol{\uplambda}{\mathord}{myletters}{"15}
\DeclareMathSymbol{\upxi}{\mathord}{myletters}{"18}

\usepackage{amsfonts}
\usepackage{lmodern} 
\usepackage{amsmath}                                                    
\usepackage{amsthm}                                                     
\usepackage{amssymb}
\usepackage{multirow}
\numberwithin{equation}{section} 
\newcommand{\newc}{\newcommand}
\newc{\be}{\begin{equation}}
\newc{\ee}{\end{equation}}
\newc{\bg}{\begin{gathered}}
\newc{\eg}{\end{gathered}}
\newc{\tref}[1]{Table \ref{#1}}
\newc{\eref}[1]{Equation \eqref{#1}}
\newc{\su}[1]{$SU(#1)$}
\newc{\fref}[1]{Figure \ref{#1}}

\newc{\ra}{\rightarrow}
\newc{\lra}{\leftrightarrow}
\newc{\ov}{\overline}
\newc{\ba}{\begin{eqnarray}}
\newc{\ea}{\end{eqnarray}}
\newc{\mf}{\mathsf}

\def\beq{\begin{equation}}
\def\eeq{\end{equation}}
\def\bea{\begin{eqnarray}}
\def\eea{\end{eqnarray}}
\usepackage{hyperref}

\begin{document}

\begin{titlepage}
\pagestyle{empty}


\vspace*{0.2in}
\begin{center}
{\Large \bf    Supersymmetric Hybrid Inflation in Light of CMB Experiments and Swampland Conjectures  }\\
\vspace{1cm}
{\bf  Waqas Ahmed $^{a,}$\footnote{E-mail: waqasmit@hbpu.ac.cn}},
{\bf  Shabbar Raza$^{b,}$\footnote{E-mail: shabbar.raza@fuuast.edu.pk}}
\vspace{0.5cm}

{\it
$^a$ Center for Fundamental Physics and School of Mathematics and Physics, Hubei Polytechnic University, Huangshi 435003, China\\
$^b$Department of Physics, Federal Urdu University of Arts, Science and Technology, Karachi 75300, Pakistan \\
}
\end{center}

\begin{abstract}
\noindent

\end{abstract}
This study revisits supersymmetric (SUSY) hybrid inflation in light of CMB experiments and swampland conjectures. We first show that if one adds radiative, soft mass, and SUGRA  corrections to the scalar potential, supersymmetric hybrid inflation is still consistent with Planck 2018 despite an impression that it does not. Usually, in SUSY hybrid inflation with minimal K\"ahler potential, the gauge symmetry breaking scale $M$ turns out to be ${\cal O}(10^{15})$ GeV, which causes proton decay rate problem. In this study, we present a new parameter space where the proton decay rate problem can be avoided by achieving $M$ of the order of $10^{16}$ GeV with $M_{S}^{2}<$0 and $am_{3/2}>$0. In this scenario, one requires a soft SUSY breaking scale $|M_{S}| \gtrsim 10^{6}$ GeV. Moreover, the tensor to scalar ratio $r$ is in the range $10^{-16}$ to $10^{-6}$, which is quite small. In this case, modified swampland hold, but it is difficult to satisfy trans-Planckian censorship conjecture. For this reason, we also consider non-minimal K\"ahler potential. We fixed spectral index $n_{S}=$0.9665 (central value) of Planck 2018 data and $M=2\times 10^{16}$ GeV and present our calculations. We show the canonical measure of primordial gravity waves  $r$ for $M_{S}=$ 1 TeV, $m_{3/2}=$ 1 TeV, $\kappa_{S}<0$ for $\cal{N}=$1 and $\cal{N}=$2, ranges from $10^{-5}$ to $0.01$ which can be observed in Planck and next-generation experiments such as LiteBIRD, Simons Observatory, PRISM, PIXIE,CORE, CMB-S4 and CMB-HD experiments that are gearing up to measure it. In addition to it, we present the parametric space and benchmark points for a non-minimal case which is consistent with modified swampland and trans-Planckian censorship conjectures. 
\end{titlepage}

\section{Introduction }
In the current era of TeV-scale physics, we have found only Higgs at the Large Hadron Collider (LHC) \cite{Aad:2012tfa,CMS} which also completes the standard model of particle physics. We have experimental and theoretical reasons to believe that there exists physics beyond the standard model (BSM). Despite the intensive searches for BSM at the LHC \cite{Aaboud:2017vwy,Sirunyan:2017cwe,Sirunyan:2017kqq,Aprile:2018dbl},  we have not found any evidence of it. Human curiosity is not just confined to the Earthbound experiments, there are satellite-based experiments such as WMAP \cite{Hinshaw:2012aka}, and Planck \cite{Akrami:2018odb} which can give glimpses on the beginning of our universe. In most of the BSM scenarios, we need additional particles besides the SM once to explain some of the mysteries of our universe. 
Supersymmetry (SUSY) is one the best extension of the SM. The minimal SUSY version of the SM that is MSSM provides the grand unification that is the gauge coupling unification of the electromagnetic, weak, and strong interactions solve gauge hierarchy problem and predict dark matter particle. The grand unification is one of the holy grail of particle physics and, since its inception, has been waiting for experimental verification such as in proton decay experiments \cite{Kearns}. It is believed that the cosmic inflation took place near or at the grand unification scale $M_{GUT}$. The cosmic inflationary models naturally demand input from particle physics. 
The cosmic inflationary models based on SUSY are very attractive. In particular, models based on SUSY grand unified theories (SUSY GUTs) provide alternative grounds to look for BSM. SUSY hybrid inflationary models \cite{Dvali:1994ms,Copeland:1994vg,Linde:1993cn,Dvali:1997uq} are very attractive because they naturally connect \cite{Chamseddine:1982jx} low-scale ($\sim$~TeV) SUSY and minimal ($N=1$) supergravity (SUGRA) \cite{Chamseddine:1982jx}.

The potential along the inflationary path at the tree level is flat in the simplest SUSY hybrid inflationary models. The radiative corrections (RC) when added to the scalar potential not only produce the required slope of the inflationary track towards the SUSY vaccum but in this process gauge symmetry group $G$ breaks to its subgroup $H$ during the inflation \cite{Jeannerot:2000sv} or at the end of the inflation \cite{Dvali:1994ms,Copeland:1994vg}.
 In such a scenario~\cite{Dvali:1994ms}, temperature anisotropy $\delta T/T$ of the cosmic microwave background (CMB) turns out to be on the order of $(M/m_P)^2$, where $M$ is the symmetry breaking scale of $G$ and $m_P$ is the reduced Planck mass.  Then, in order to keep intact the self-consistency of the inflationary scenario $M$ is comparable to the scale of grand unification, $M_{GUT} \sim 10^{16}$~GeV, hinting that $G$ may constitute a grand unified theory (GUT) group. 

\textcolor{black}{In SUSY hybrid inflation, it is important to note that the scalar spectral index $n_{\rm s}$ falls within the observed range of 0.96-0.97, as indicated by \cite{Ade:2015lrj}. This alignment is achievable when the inflationary potential incorporates radiative corrections \cite{Dvali:1994ms} along with either the inclusion of soft SUSY breaking terms \cite{rehman,gravitywaves} or higher-order terms in the K\"{a}hler potential \cite{bastero}.
} On the other hand, if such terms are not included,
	the $n_{\rm s}$ lies close to 0.98. Such a value of $n_{s}$ is only acceptable if the effective 
	number of light neutrino species is slightly higher than 3 \cite{Ade:2015lrj}.
	Since one requires 50 or so e-foldings to resolve the horizon and flatness problem, so the inflaton field values remain well below the Planck scale $M_{P}$, so in both cases, the supergravity corrections remain under control.

In this article, we aim to tackle two issues. In 2018, the Planck collaboration \cite{Akrami:2018odb}, and last year's publication from LiteBIRD \cite{LiteBIRD:2022cnt}, highlighted models that still hold promise for observation in future experiments. In Figure 8 of \cite{Akrami:2018odb} and Figures 2 and 3 of \cite{LiteBIRD:2022cnt}, it is asserted that the CMB predictions from \cite{Dvali:1994ms} appear to have been ruled out.  We address this problem in two ways. In the first step, we employ the minimal K\"{a}hler potential, and taking into account radiative corrections, soft SUSY contributions along with SUGRA corrections, and show that SUSY hybrid inflation is consistent with the Planck 2018 \cite{Akrami:2018odb} but suffers one problem. The problem is that the gauge symmetry breaking scale in these scenarios comes out to be ${\mathcal O}(10^{15})$ GeV, which causes the proton decay rate problem. One can avoid this problem if $M$ is ${\mathcal O}(10^{16})$ GeV. We display a parameter space where we not only have $n_{s}$ consistent with Planck 2018  data but also achieve $M$ of order $10^{16}$ GeV and thus avoid the proton decay rate problem. In these calculations, we use ${\mathcal N}=$1 and ${\mathcal N}=$2, number of e-folding $N_{0}$=50 and show that with $M_{S}^{2}<$0 and $am_{3/2}>$0 we can achieve this task. Here we have $m_{3/2}\neq|M_{S}|$ and $m_{3/2}=$1,10,100 and 1000 TeV. The price we have to pay is that soft SUSY breaking scale $|M_{S}|\gtrsim 10^{6}$ GeV and scalar to tensor ratio $r$ is very small (between $10^{-15}$ to $10^{-8}$). Since in this scenario, $M_{S}^{2}<$0 and values of $r$ is very tiny, we also employ non-minimal K\"ahler potential. In this case we chose $M=2\times 10^{16}$ GeV, $N_{0}=$50, $M_{S}=am_{3/2}=$1 TeV with $a=-1$ along with $\kappa_{S}<$0. We show that in this case $r$ can be in the range $3\times 10^{-5}$ to $0.01$ while $|S_{0}|/m_{p}$ is in the range $0.02$ to $1$. 

The second issue we address in this article is the proposed swampland conjectures \cite{Agrawal:2018own,Obied:2018sgi,Ooguri:2018wrx,Andriot:2018mav} (we will discuss it in detail later in this article). {It is believed that cosmic inflation took place at or near $M_{GUT}$  and one may describe it by the low-energy  effective field  theories(EFTs) \cite{Guth:1980zm,Ahmed:2018jlv,Schmitz:2019uti}.  Such EFTs may only be trusted if we can embedded them successfully in a quantum theory of gravity such as string theory.  Such a subset of EFTs with UV completion belongs to the landscape, while the rest are said to lie in the swampland since such theories render quantum gravity inconsistent when coupled to gravity.  For a detailed review, see ref. \cite{Schmitz:2019uti}. The proposed swampland criteria has emerged from the weak gravity conjecture \cite{Montero:2018fns} and corresponding progress in string theory and black hole physics \cite{Dvali:2018jhn}. These swampland criteria can be used as consistency check on the EFTs to see if they lie in the swampland or in landscape. We briefly discuss swampland criterion and Trans-Planckian censorship conjecture (TCC) \cite{Bedroya:2019snp,Bedroya:2019tba} in hybrid inflationary models.

 The rest of the paper is organized as follows. In section \ref{sec2} we give detail of the SUSY hybrid inflation model with minimal K\"ahler potential. In section \ref{sec3} we describe swampland conjecture and its variations. We present our numerical calculations of SUSY hybrid inflation with minimal and non-minimal  K\"ahler potential in section \ref{sec4} . Finally, we summarize our results in section \ref{sec6}.

\section{\large{\bf Supersymmetric Hybrid Inflation}}\label{sec2}

SUSY hybrid inflation models are described by the following normalizable superpotential \cite{Dvali:1994ms,Copeland:1994vg}
\begin{equation} 
W=\kappa S(\Phi \overline{\Phi }-M^{2})\,,
\label{superpot}
\end{equation}
where $S$ is a gauge singlet superfield representing the inflaton, and $\Phi$, $\overline{\Phi }$ are conjugate superfields which transform nontrivially under some gauge group $G$, and provide the vacuum energy associated with inflation, $\kappa$ is a dimensionless coupling constant and $M$ is mass parameter correspond to the scale at which gauge group $G$ is broken. In addition to it, the superpotential given in Eq.~\ref{superpot} possesses the gauge symmetry $G$ along with an additional $U(1)$ R-symmetry which is usually employed to suppress proton decay if $Z_{2}$ is its subgroup. Also, $U(1)$ R-symmetry provides a unique structure of superpotential, linear in S.  The global SUSY F-terms are given by,
\begin{equation}\label{VF1}
V_{F}=\sum_{i}|\frac{\partial W}{\partial z_{i}}|^{2},
\end{equation}
where $z_{i}\in \{\phi , \overline{\phi }, s\}$  represents the bosonic components of the superfields $\Phi, \overline{\Phi },S$ respectively. Using Eqs.~(\ref{superpot})--(\ref{VF1}), one can write the tree level global SUSY potential in the D-flat direction $|\phi|=|\overline{\phi}|$ as

\begin{equation}\label{eq1}
V= \kappa^2\,(M^2 - \vert \phi\vert^2)^2 + 2\kappa^2 \vert s \vert^2 \vert \phi \vert^2 .
\end{equation}

\textcolor{black}{Inflation initiates along the local minimum at $|\phi| = 0$ (the inflationary track), commencing from a large $|s|$. An instability emerges at the waterfall point, denoted as $|s_c|^2 = M^2$, where $s_c$ is the value of $|s|$ at which $\frac{\partial^2V}{\partial s^2}\mid_s=0$. At this point, the field naturally transitions into one of the two global minima at $|\phi|^2=M^2$, coinciding with the breaking of the gauge group G. The scalar potential is approximately quadratic in $|\phi|$ at large $|s|$, while at $|s|=0$, equation \ref{eq1} transforms into a Higgs potential.}

Along the inflationary track the gauge group $G$ is unbroken, however, only the constant term $V_0=\kappa^2 M^4$ is present at tree level, thus SUSY breaks during inflation. When SUSY breaks it splits fermionic and bosonic mass multiplets and it contributes to radiative corrections. The minimal (canonical) K\"ahler potential may be written as
\begin{equation}
K=  |S|^{2}+ |\Phi|^{2} + |\overline{\Phi}|^{2}.  \label{kahler}
\end{equation}
The SUGRA scalar potential is given by
\begin{equation}
V_{F}=e^{K/m_{P}^{2}}\left(
K_{ij}^{-1}D_{z_{i}}WD_{z^{*}_j}W^{*}-3m_{P}^{-2}\left| W\right| ^{2}\right),
\label{VF}
\end{equation}
where $z%
_{i}\in \{\phi , \overline{\phi }, s,\cdots\}$ represents the bosonic components of the superfields, and here we have also defined
\begin{equation*}
D_{z_{i}}W \equiv \frac{\partial W}{\partial z_{i}}+m_{P}^{-2}\frac{\partial K}{\partial z_{i}}W, \,\,\,\,\,\, K_{ij} \equiv \frac{\partial ^{2}K}{\partial z_{i}\partial z_{j}^{*}},
\end{equation*}
$D_{z_{i}^{*}}W^{*}=\left( D_{z_{i}}W\right) ^{*}$ and $m_{P}=M_P/\sqrt{8\pi} \simeq 2.4\times 10^{18}$~GeV is the reduced Planck mass. By including the various term such as leading order SUGRA corrections, radiative correction and soft SUSY breaking terms, we may write the scalar potential along the inflationary trajectory (i.e. $|\phi|=|\overline{\phi}|=0$) as,
\begin{eqnarray}
V &\simeq& V_{\text{SUGRA}} + \Delta V_{\text{1-loop}} + \Delta V_{\text{Soft}}, \\
&\simeq&
\kappa ^{2}M^{4}\left( 1 + \left( \frac{M}{m_{P}}\right) ^{4}\frac{x^{4}}{2}+\frac{\kappa ^{2}\mathcal{N}}{8\pi ^{2}}F(x) + a\left(\frac{m_{3/2}\,x}{\kappa\,M}\right) + \left( \frac{M_S\,x}{\kappa\,M}\right)^2\right).
\label{scalarpot}
\end{eqnarray}
Here 
\begin{equation}
F(x)=\frac{1}{4}\left( \left( x^{4}+1\right) \ln \frac{\left( x^{4}-1\right)}{x^{4}}+2x^{2}\ln \frac{x^{2}+1}{x^{2}-1}+2\ln \frac{\kappa ^{2}M^{2}x^{2}}{Q^{2}}-3\right)
\end{equation}
represents radiative corrections, and 
\begin{equation}
a = 2\left| 2-A\right| \cos [\arg s+\arg (2-A)].
\label{a}
\end{equation}
Moreover, we  the inflaton field by $x\equiv |s|/M$,  the dimensionality of the representation of the conjugate pair of fields $\phi$ ($\overline{\phi}$) with respect to the gauge group G is given by $\mathcal{N}$, $m_{3/2}$ is the gravitino mass, and $Q$ is the renormalization scale.  In this article, we employ $G\equiv U(1)$ which may be identified as $U(1)_{B-L}$ corresponding to $\mathcal N$=1 and the left-right symmetric model $G\equiv SU(3)_{C}\times SU(2)_{L}\times SU(2)_{R}\times U(1)_{B-L}$}  corresponds to $\mathcal N=$2. In Eq.~(\ref{scalarpot}) the last two term correspond to soft SUSY-breaking linear and mass-squared terms, respectively and can be derived from a gravity mediated SUSY breaking scheme \cite{Dvali:1997uq}.


Using the standard slow-roll approximation, various inflationary parameters can estimated which are given below:

\bea
\epsilon = \frac{1}{4}\left( \frac{m_P}{M}\right)^2
\left( \frac{V'}{V}\right)^2, \,\,\,
\eta = \frac{1}{2}\left( \frac{m_P}{M}\right)^2
\left( \frac{V''}{V} \right), \,\,\,
\xi^2 = \frac{1}{4}\left( \frac{m_P}{M}\right)^4
\left( \frac{V' V'''}{V^2}\right). 
\label{slowroll}
\eea

In addition to this, in the slow-roll approximation, the scalar spectral index $n_s$, the tensor-to-scalar ratio $r$ and the running of the scalar spectral index $dn_s / d \ln k$ are given by
\bea
n_s &\simeq& 1+2\,\eta-6\,\epsilon,  \\ \label{ns}
r &\simeq& 16\,\epsilon, \label{r0} \\
\frac{d n_s}{d\ln k} &\simeq& 16\,\epsilon\,\eta
-24\,\epsilon^2 - 2\,\xi^2 .
\eea

The Planck constraint on the scalar spectral index $n_{s}$ in the $\Lambda$CDM model for very small values of $r$ and $\frac{d n_s}{d\ln k}$ is\cite{Akrami:2018odb}
\be                                         
n_s = 0.9665 \pm 0.0038  \,\,\,\,\,\,\, (68\% CL, Planck\,TT,TE,EE+lowE+lensing+BAO).
\ee
The amplitude of the primordial spectrum is given by,
\be
A_{s}(k_0) = \frac{1}{24\,\pi^2}
\left. \left( \frac{V/m_P^4}{\epsilon}\right)\right|_{x = x_0},  \label{curv}
\ee
and has been measured by Planck to be $A_{s}= 2.137 \times 10^{-9}$ at $k_0 = 0.05\, \rm{Mpc}^{-1}$ \cite{Ade:2015lrj}.
The last $N_0$ number of e-folds before the end of inflation is,
\bea
N_0 = 2\left( \frac{M}{m_P}\right) ^{2}\int_{x_e}^{x_{0}}\left( \frac{V}{%
	V'}\right) dx,
\eea
here $x_e$ is the value of the field at the end of inflation and $x_0$ define as the value of the field at the pivot scale $k_0$. 
\textcolor{black}{The value of $x_e$ is constrained either by the breakdown of the slow roll approximation or by a 'waterfall' destabilization event taking place at the critical value $x_c = 1$, if the slow roll approximation holds}.

\section{Swampland Conjecture and its variation}\label{sec3}
It is believed that cosmic inflation took place at or around $M_{GUT}$  and one may describe it by the low-energy effective field  theories(EFTs) \cite{Guth:1980zm,Ahmed:2018jlv,Schmitz:2019uti}. If such theories have anything to do with the possible theories that may describe the observed universe, these EFTs could be successfully embedded in quantum theory of  gravity such as string theory. The subset of such EFTs with UV completion are said to belong to the landscape while rest of the EFTs lie in the swampland. The EFTs lying in swampland when couple to gravity render quantum gravity inconsistent. Recent development in quantum gravity and related physics \cite{Dvali:2018jhn}   
lead to the  proposal of two swampland criteria that can be employed as a consistency check on the EFT that ensures it to lie in the landscapes. The swampland conditions \cite{Agrawal:2018own, Obied:2018sgi} relevant for inflation are as follows:

\begin{itemize}
	\item \emph{Distance conjecture}. Distance conjecture restricts scalar field excursions in reduced Planck units
	$(m_p=\sqrt{\hbar c/8\pi G}=1)$ such that,
	\bea
	\fbox{$ \Delta \phi < 1 $} \label{eq:distance}
	\eea
	Whenever $\Delta\phi$ exceeds 1, an infinite tower of states become exponentially light and the effective field theory description breaks down \cite{Agrawal:2018own}.
	\item \emph{Refined de~Sitter conjecture}. In \cite{Obied:2018sgi} de~Sitter conjecture was first proposed that forbade both de~Sitter maxima and de~Sitter minima. Later, when a number of counter examples for de~Sitter maxima were reported, redefined de~Sitter conjecture allowing the maxima but still forbidding the de~Sitter minima has been argued \cite{Ooguri:2018wrx},
	\bea
	|\nabla V| \geq c\,V \quad {\rm or}  \quad  \min(\nabla_i \nabla_j V)  \leq  -c' V \,, \label{V-RdS}
	\eea
	where $c,c'$ are constants of $\mathcal{O}(1)$.
	The above relations have been combined into one relation in \cite{Andriot:2018mav}:
	\beq
	\left(\frac{|\nabla V|}{V}\right)^q - a\,\frac{{\rm min}\nabla \partial V}{V} \geq b\,, \qquad  {\rm with}\ a+b=1, \quad a,b>0, \quad q > 2  \quad \label{V-RdSconj}
	\eeq
	Defining $c\equiv V'/V$ and $c'\equiv-V''/V$, we can re-write it as,
	\beq
	\fbox{ $ c^q + a\, (c'+1) \geq 1,  \quad   0<a<1, \, q>0  $} \label{V-RdSC}
	\eeq
\end{itemize}
\begin{itemize}
{\item \emph{
	Trans-Planckian censorship conjecture
	(TCC).} Bedroya and Vafa proposed another conjecture name trans-Planckian censorship conjecture (TCC)\cite{Schmitz:2019uti,Bedroya:2019snp,Bedroya:2019tba}, which says that any
	inflation model that stretches modes on sub-Planckian
	length scales to the extent that they eventually exit the Hubble horizon must lie in the swampland. The TCC has important implications for inflationary cosmology.  In particular, it
	severely constrains the allowed range for the total number of e-folds $N_{0}$
	during inflation, which translates into upper bounds on the scale of inflation $H$ and tensor-to-scalar ratio $r$}.

\end{itemize}
These developments have generated a flurry of papers to ascertain the effects of swampland conditions on inflationary models \cite{Schmitz:2019uti}. These conjectures have severely constrained the inflationary models in 4-dimensions. After the Vafa etal conjecture many physicist try to address the swampland conditions specially the conditions related to de Sitter conjectures because its contradict the experimental results. Dvali etal~\cite{Dvali:2018jhn} explain the de sitter condition in terms of quantum breaking in the following way;.
\begin{itemize}
 \item \emph{ Bound on Extrema.}
 The classical break-time $t_{CL}$ is defined as the time-scale after which the classical nonlinearities fully change the time evolution described by a free system. On the other hand, the quantum break-time $t_Q$ is the time-scale after which the system can no longer be studied classically, no matter how well 
one accounts for classical nonlinearities. It has been shown by Dvali etal~\cite{Dvali:2018jhn} that 
for a system with generic interaction strength say $\kappa$, the quantum break-time is given 
by $t_Q = t_{Cl}/\kappa$. To avoid quantum de-Sitter break-time implies the following bound
 
    \bea
    V''&\lesssim &-V,\label{eq:t_QB} \quad \rm\emph{Quantum~Breaking~Bound}
    \eea

Recall that one of the slow parameter $\eta\equiv \frac{V''}{V}<< 1$ which implies that the above mentioned quantum breaking-bound is equivalent to $\eta$.

 \item \emph{Bound on Slow-Roll.}
 In slow-roll regime,  the potential $V(\phi)$ is considered away from extrema. Due to the quantum breaking bound, the change of the potential $\Delta V$ over some time $\Delta t \sim t_Q$ must satisfy $|\Delta V| \gtrsim V$. One can write approximately $\Delta V \sim V'\dot{\phi}\Delta t$, If we demand the slow-roll criteria to be satisfied, we can write $\dot{\phi}\sim - V'/H$, which implies $\Delta V \sim -V'^2\Delta t/H$. Since we are considering the slow-roll case, one can write $t_{Cl} \sim H^{-1}$. Finally, we obtain
 the following bound whenever $V>0$:
    \bea
    \frac{|V^\prime|}{V}&\gtrsim & \sqrt{\alpha}, \quad \rm\emph{de~Sitter~Conjecture}\label{D-dSC}
    \eea
    where $\alpha= N_{sp}H^2$ is an effective strength of graviton scattering for the characteristic momentum transfer $H$,
 and $N_{sp}$ is the number of light particle species~\cite{Dvali:2017eba}. Note that another slow-roll parameter $\epsilon$ can be written as $\sqrt{2\epsilon}\equiv\frac{|V^\prime|}{V}$. This implies that $\sqrt{2\epsilon} \gtrsim \sqrt{\alpha}$.
\end{itemize}

\section{Numerical Analysis}{\label{sec4}}
In this section, by using slow roll equations~\ref{slowroll}, we do a series of numerical calculations to analyze the implications of the model and discuss its predictions regarding the various cosmological observables. We divide our analysis into two cases, that is with minimal K\"ahler and non-minimal K\"ahler potential.  We show that the SUSY hybrid inflation model with minimal K\"ahler potential is still consistent with Planck 2018 without proton decay problem and modified swampland conjectures. However, satisfying TCC  conjecture is difficult in minimal case.  To address, large $r$ solutions and TCC conjecture,  we consider Hybrid inflation with non-minimal K\"ahler.  With non-minimal K\"ahler potential, we have large r as well as very tiny r, which is required to satisfied swampland and TCC conjectures. The details of numerical results are presented in the following sub sections.
\begin{figure}[t!]
	
	\begin{subfigure}{.5\textwidth}
		\centering	
		\includegraphics[width=1\linewidth]{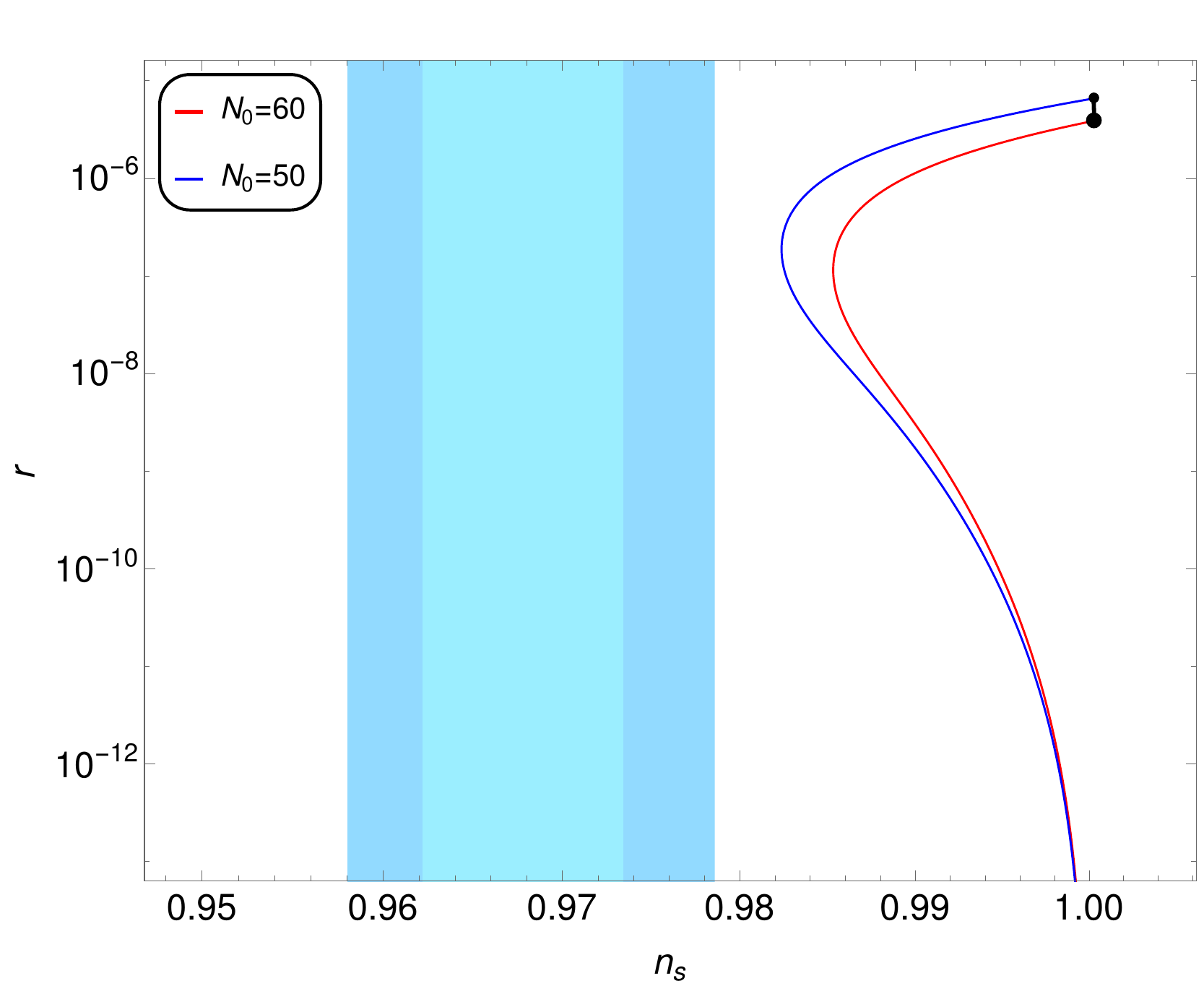}
	\end{subfigure}
	\begin{subfigure}{.5\textwidth}
		\centering	
		\includegraphics[width=1\linewidth]{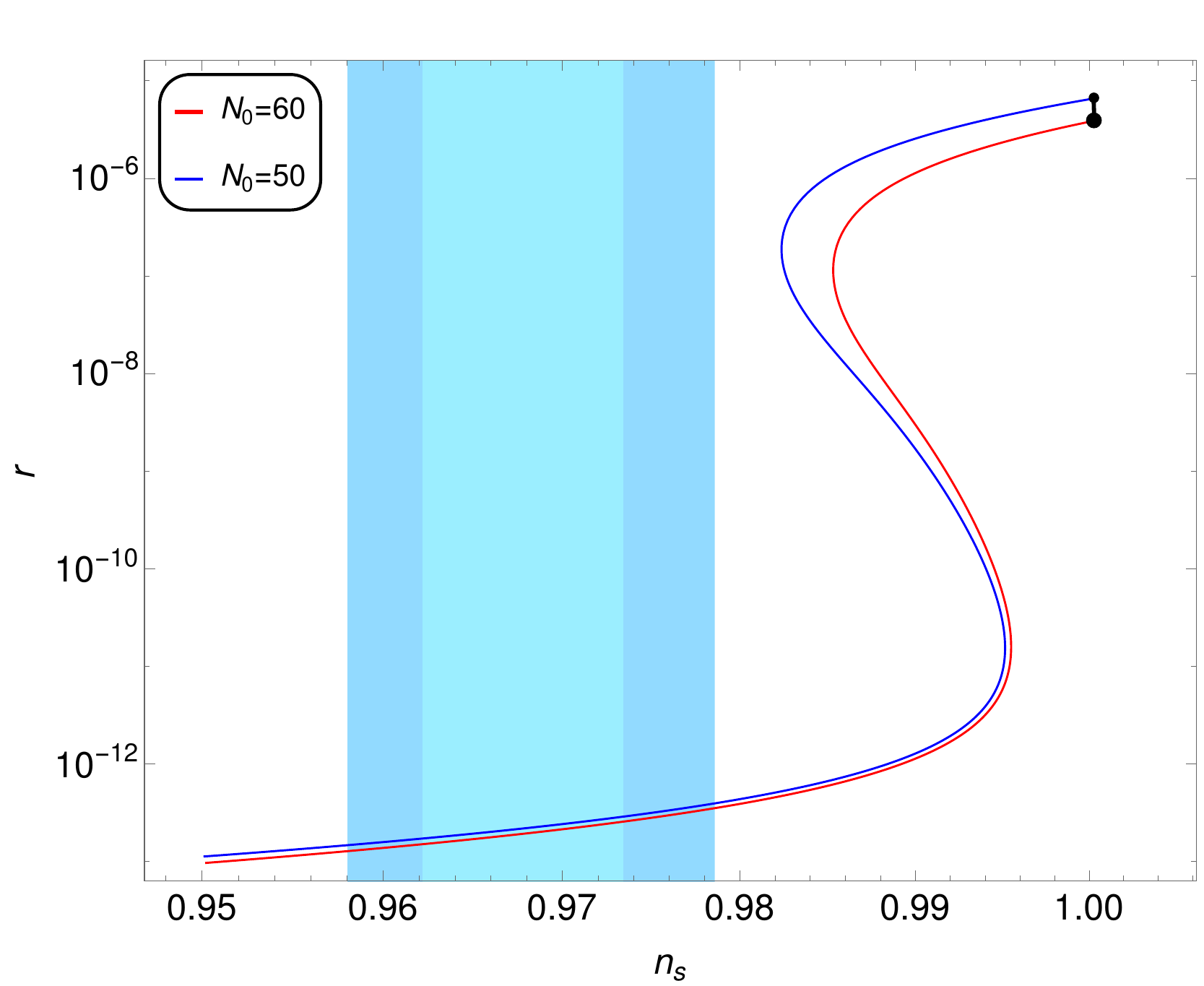}
	\end{subfigure}
	\caption{\small{Figure shows plots for  $n_{s}$ v.s $r$ for ${N_{0}}=$50 (blue curve) and ${N_{0}}$=60 (red curve). The vertical light blue (dark blue) color shows 1 $\sigma$ (2 $\sigma$) variation from central value of $n_s = 0.9665$ \cite{Akrami:2018odb}}.}
	\label{fig1}
\end{figure}

\subsection{SUSY Hybrid Inflation and CMB Experiments}
 In Fig.~\ref{fig1} we present plots in $r-n_{s}$ plane. In the left plot, we only include the first two terms of the scalar potential \ref{scalarpot} that is only SUGRA and radiative corrections. The plot in the right panel also includes a contribution of soft SUSY breaking terms. In these plots we put number of e-folding ${N_{0}}$=50 and  $N_{0}$=60 while keeping $M_{S}=am_{3/2}$= 1 TeV with sign of $a$ negative. The horizontal light blue (dark blue) band displays 1$\sigma$ (2$\sigma$) variation from the central values of $n_{s}$=0.9665 \cite{Akrami:2018odb}. From the left panel, it is evident that without adding contributions of soft SUSY breaking terms to the scalar potential, the hybrid-inflation model can be ruled out, according to Planck 2018. On the other hand, from the plot in the right panel, we see that if we also include contributions of soft SUSY breaking terms to the scalar potential, the hybrid-inflationary model is still consistent with \cite{Akrami:2018odb}. In this plots we also notice that to remain within 2$\sigma$ variation in $n_{s}$, $r$ should be less than $10^{-12}$.  In minimal W and K, we divide our analysis in two cases like $M^{2}_{S}>0$ and $M^{2}_{S}<0$ which we will discuss below briefly.  
 \subsubsection{$M^{2}_{S}>0$}
\begin{figure}
	
	\begin{subfigure}{.5\textwidth}
		\centering	
		\includegraphics[width=1\linewidth]{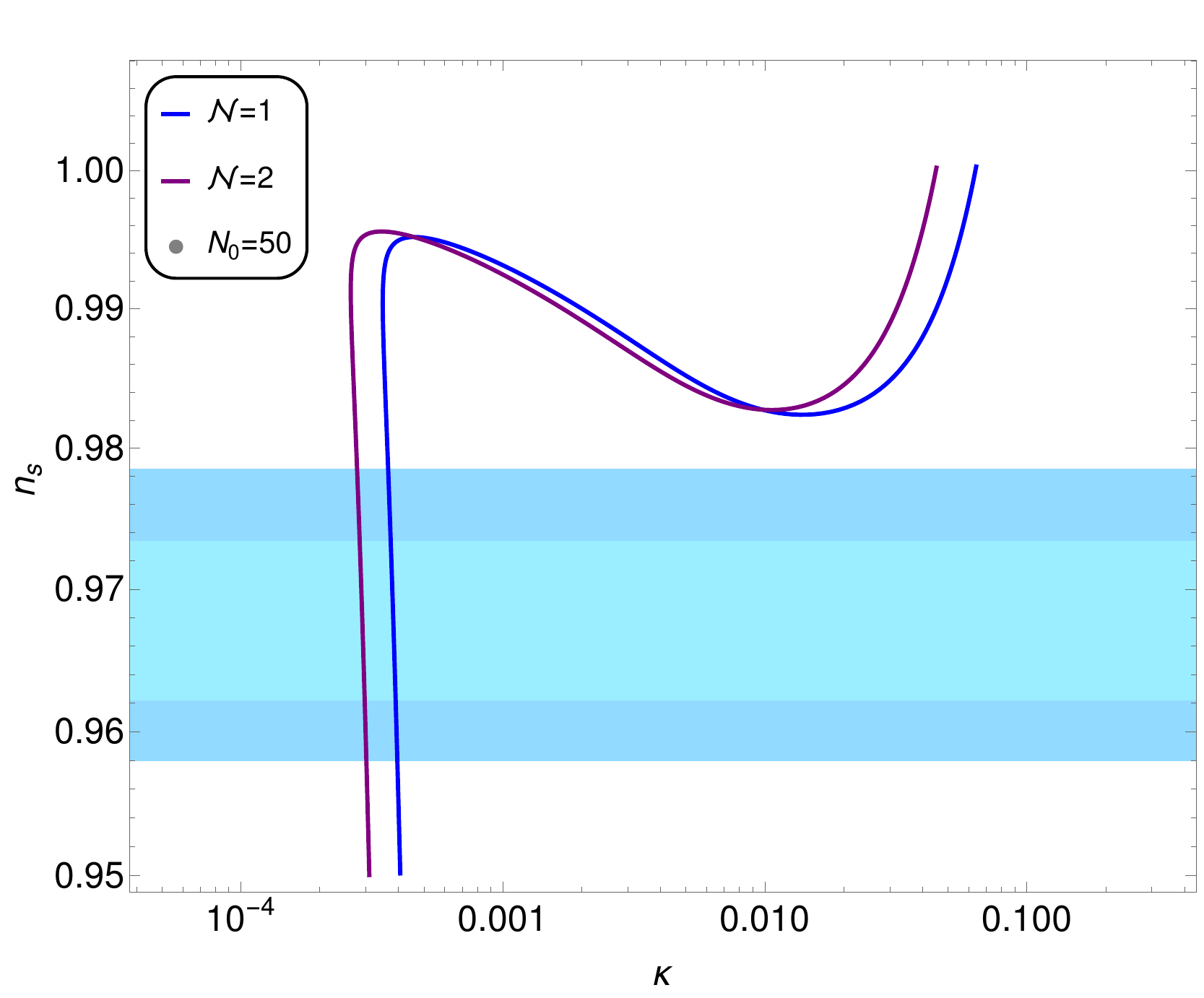}
	\end{subfigure}
	\begin{subfigure}{.5\textwidth}
		\centering	
		\includegraphics[width=1\linewidth]{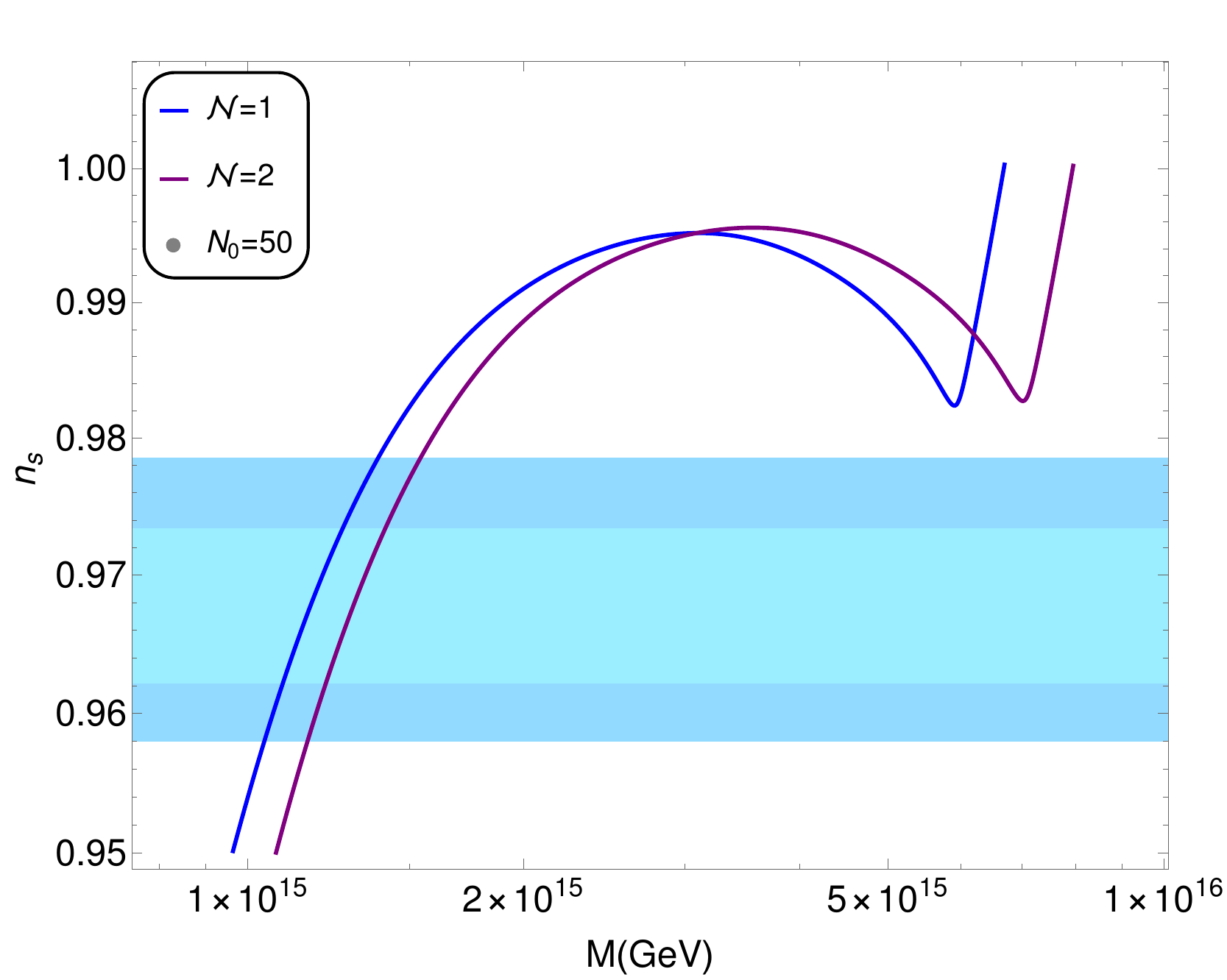}
	\end{subfigure}
	\caption{\small{Figure shows plots for  $n_{s}$ v.s $\kappa$  and $M$ v.s $\kappa$ for $\mathcal{N}=1$ (blue curves) and $\mathcal{N}=2$ (purple curves). The horizontal light blue (dark blue) color shows 1 $\sigma$ (2 $\sigma$) variation from central value of $n_s = 0.9665$ \cite{Akrami:2018odb}}.}
	\label{fig2}
\end{figure}

In Fig.~\ref{fig2}, we show plots in $\kappa-n_{s}$ and $M-n_{s}$ planes with $M_{S}=am_{3/2}$= 1 TeV with sign of $a$ negative for $\mathcal{N}$=1 (blue curve) and  $\mathcal{N}$=2 (purple curve). Horizontal blue bands have the same color coding as in Fig.~\ref{fig1}. One can see from the left panel that to remain consistent with the Plank 2018 data we have $10^{-4}\lesssim \kappa \lesssim 10^{-3}$ for both $\mathcal{N}$=1 and  $\mathcal{N}$=2. Plot in the right panel reveals that in this scenario the gauge symmetry breaking scale $M$ is in the range $\sim 1 \times 10^{15}$ to $1.5\times 10^{15}$ GeV. This small value of $M$ can cause proton-decay problem despite the fact that our results are consistent with Planck 2018 \cite{Akrami:2018odb}.

\subsubsection{$M^{2}_{S}<0$}
\begin{figure}
	\begin{subfigure}{.5\textwidth}
		\centering	
		\includegraphics[width=1\linewidth]{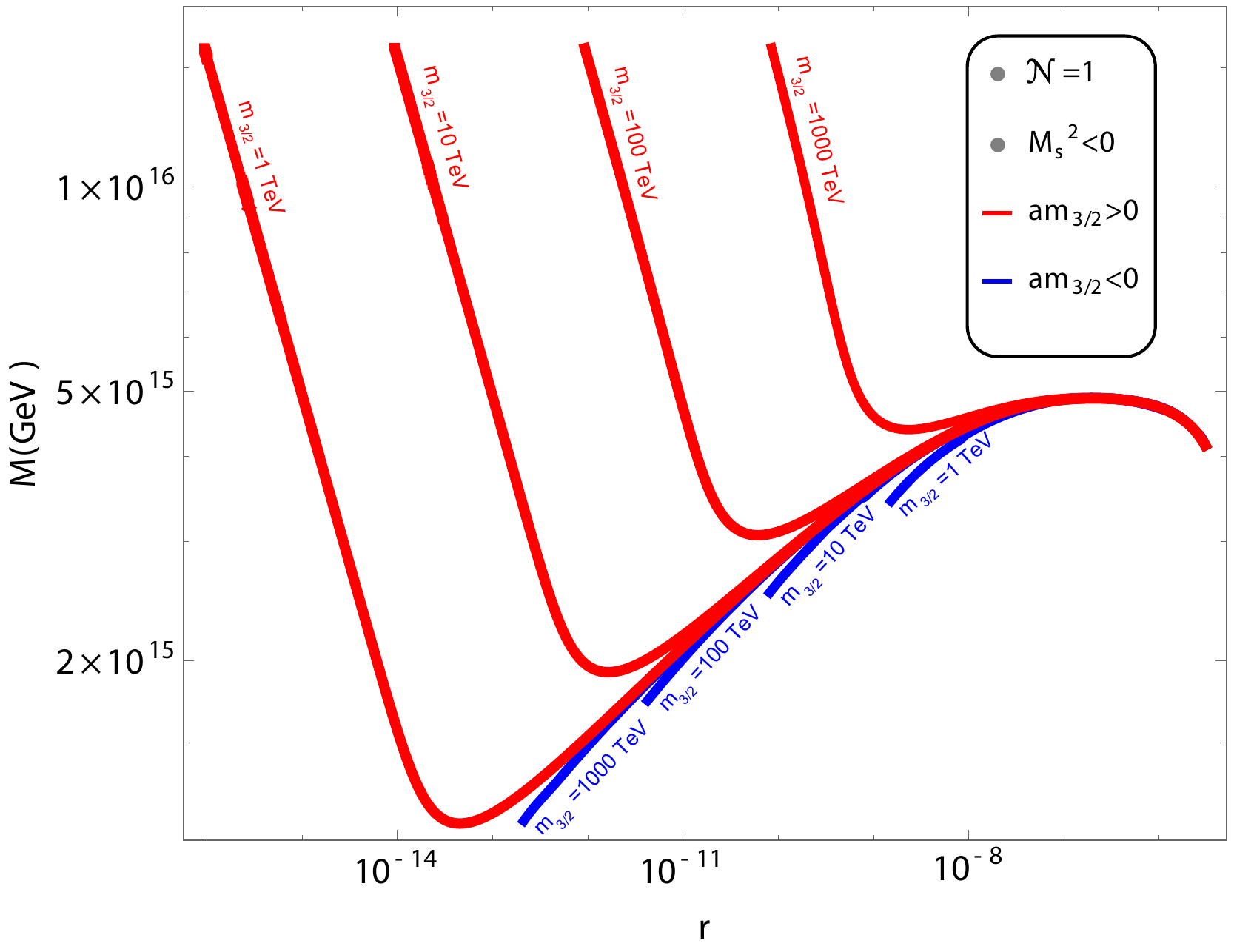}
		\label{4d2}
	\end{subfigure}
	\begin{subfigure}{.5\textwidth}
		\centering	
		\includegraphics[width=1\linewidth]{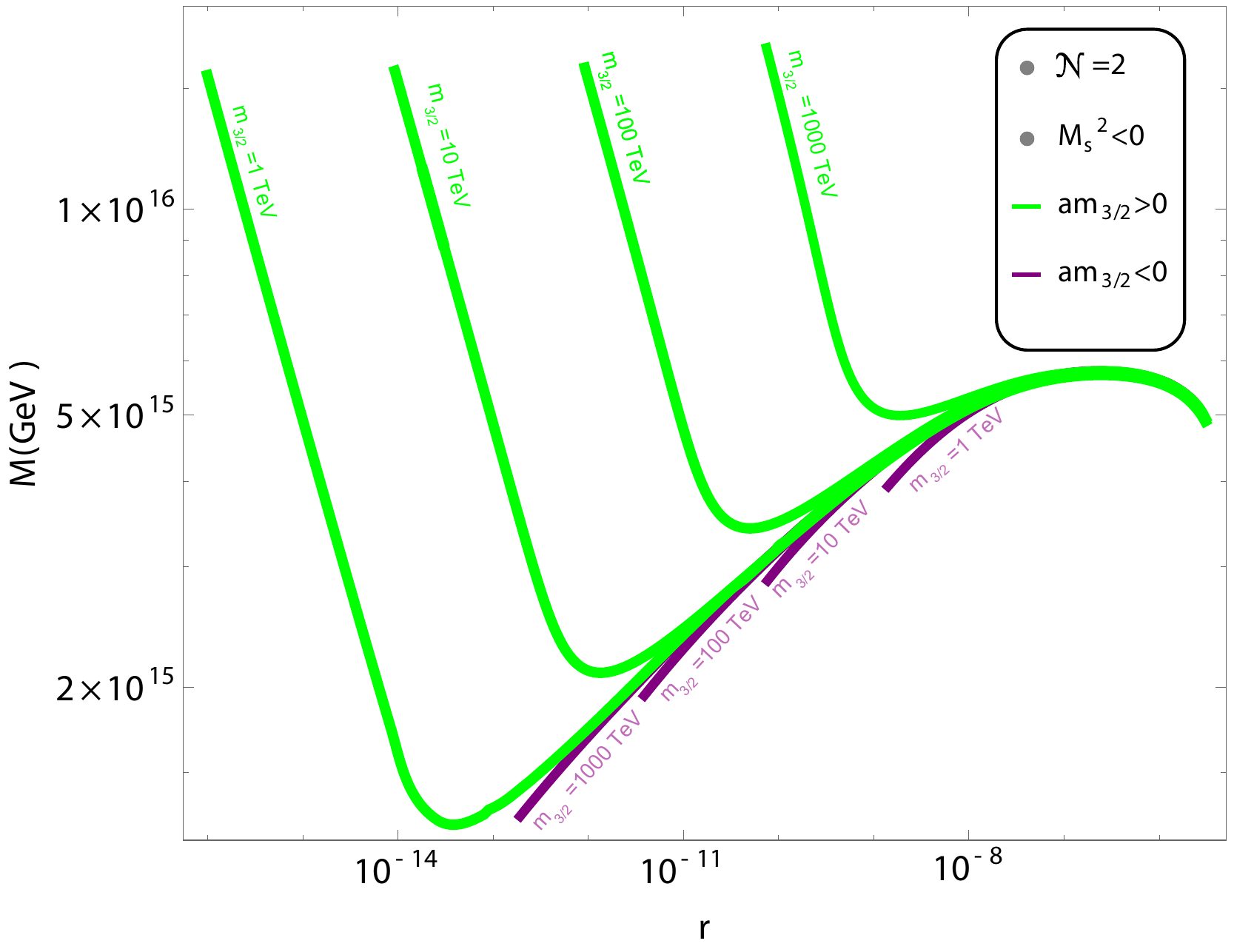}
		\label{4d2}
	\end{subfigure}
	\begin{subfigure}{.5\textwidth}
		\centering	
		\includegraphics[width=1\linewidth]{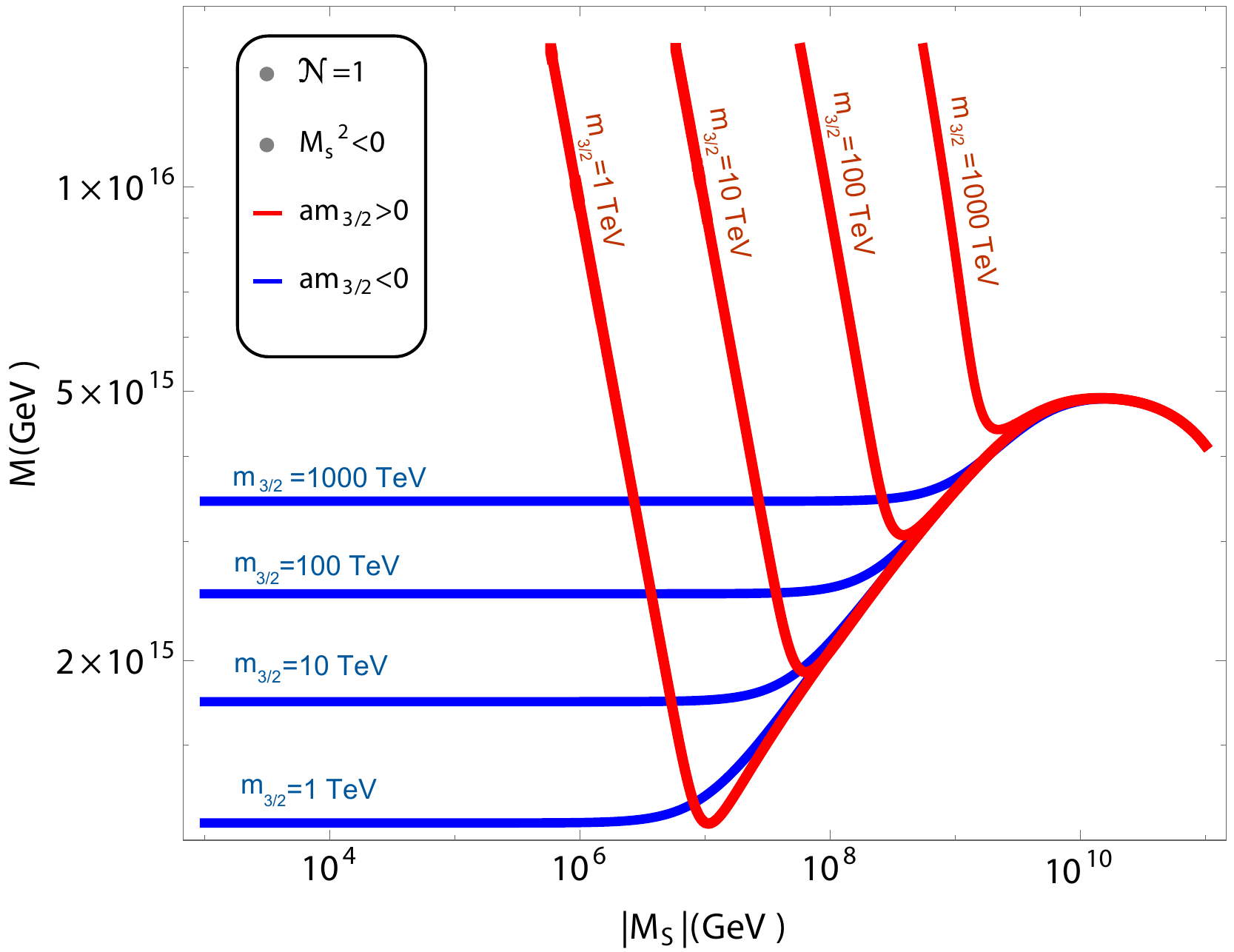}
		\label{4d2}
	\end{subfigure}
	\begin{subfigure}{.5\textwidth}
		\centering
		\includegraphics[width=1\linewidth]{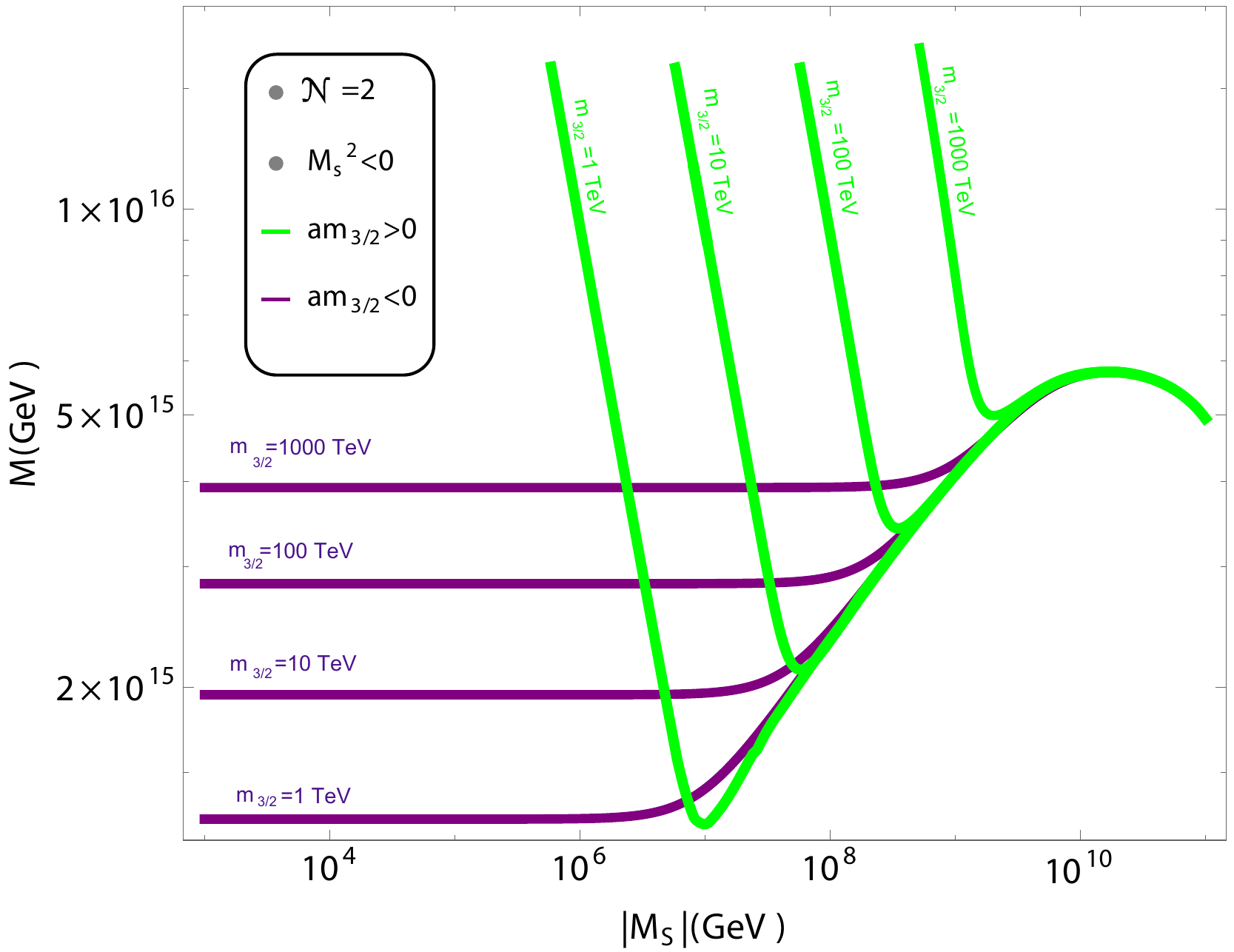}
		\label{4d2}
	\end{subfigure}\\
	\caption{\small{Plots in $r-M$  (upper panels) and $|M_{S}|-M$ (lower panel) for $N_0 =$50, $n_s = 0.9665$ (central value) \cite{Akrami:2018odb} and $M_{S}^{2}<0$. Curves for $am_{3/2} >$0 (red color) and $am_{3/2} <$0 (blue color) for $\mathcal{N}$=1, and curves for $am_{3/2} >$0 (green color) and $am_{3/2} <$0 (purple color) for $\mathcal{N}$=2 are drawn for $m_{3/2}$=1,10,100 and 1000 TeV. }}
	\label{fig3}
\end{figure}

\begin{figure}
	\begin{subfigure}{.5\textwidth}
		\centering	
		\includegraphics[width=1\linewidth]{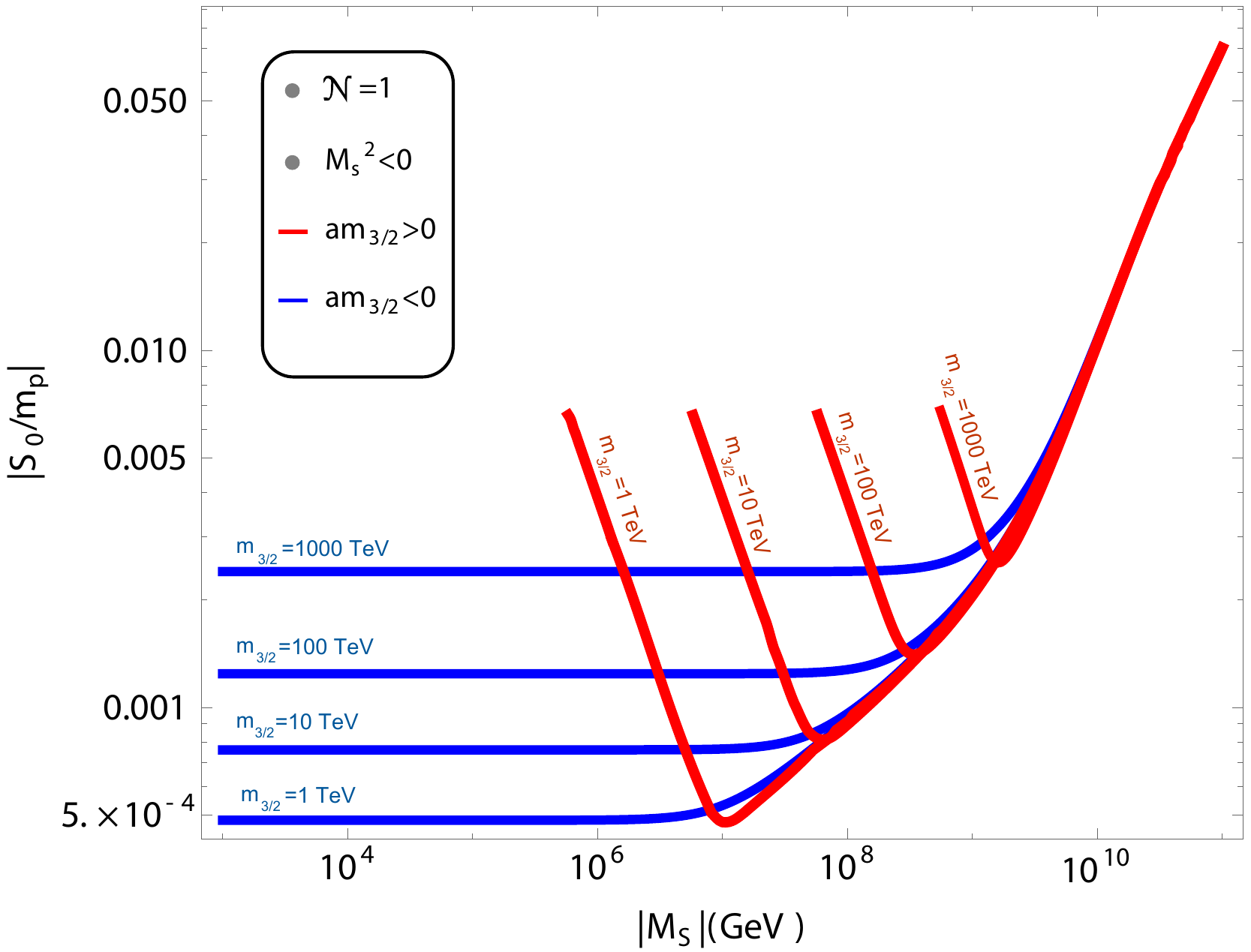}
	\end{subfigure}
	\begin{subfigure}{.5\textwidth}
		\centering	
		\includegraphics[width=1\linewidth]{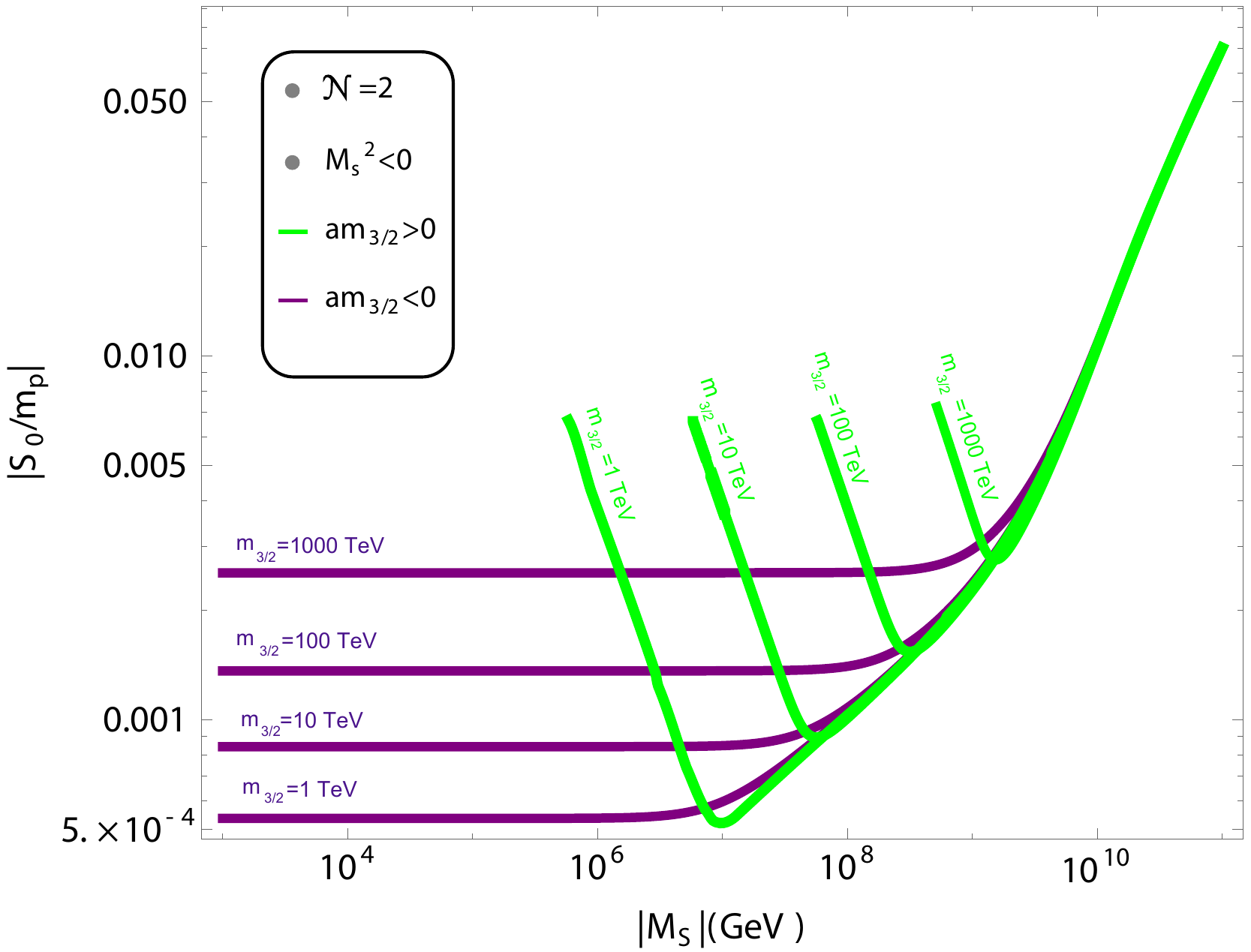}
		\label{4d2}
	\end{subfigure}\\
	\begin{subfigure}{.5\textwidth}
		\centering	
		\includegraphics[width=1\linewidth]{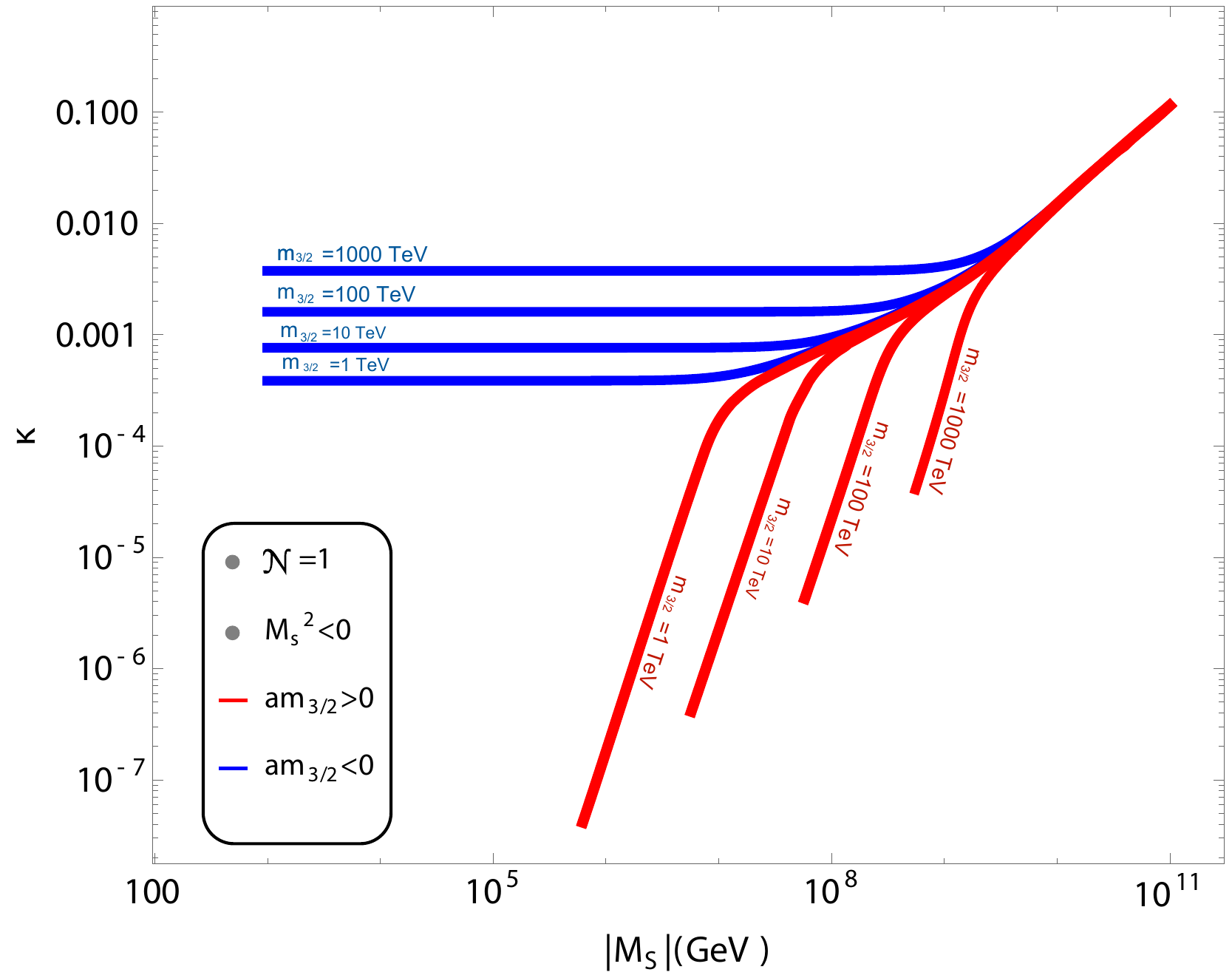}
		\label{4d2}
	\end{subfigure}
	\begin{subfigure}{.5\textwidth}
		\centering
		\includegraphics[width=1\linewidth]{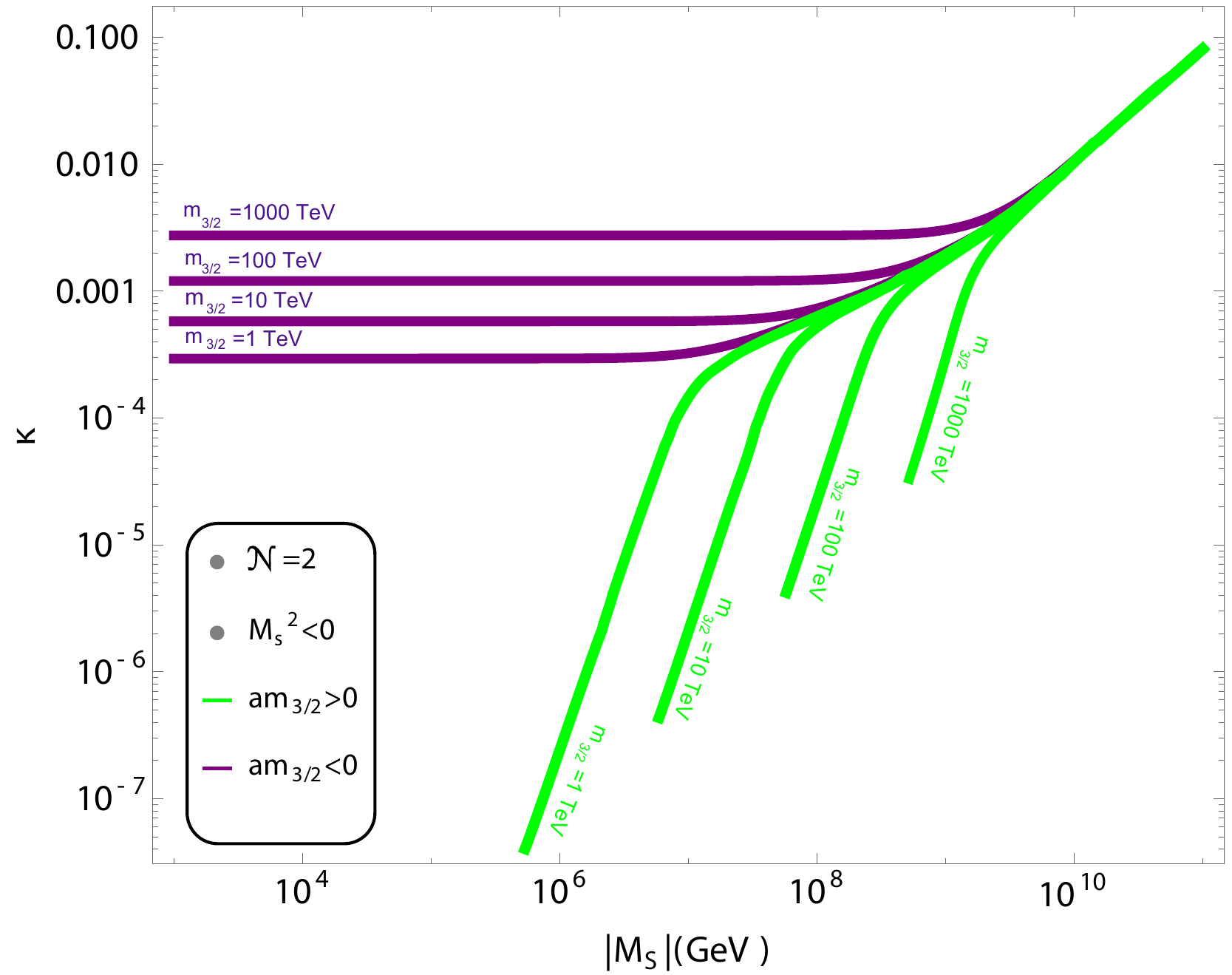}
		\label{4d2}
	\end{subfigure}	%
	\caption{\small{Plots in $|M_{S}|-|S_{0}|/m_{p}$  (upper panels) and $|M_{S}|-\kappa$  (lower panels) planes. Color coding is same as in Fig.~\ref{fig3}.}}
	\label{fig4}
\end{figure}

 In this sub section we briefly discuss how in minimal Kh\"{a}ller potential, one can overcome smallness of $M$ problem while still satisfying current Planck bounds \cite{Akrami:2018odb} on $n_{s}$. In Fig.~\ref{fig3} and Fig.~\ref{fig4} we keep $n_{s}=$0.9665 (central value) \cite{Akrami:2018odb} and $N_{0}=$50, $m_{3/2} \neq M_{S}$. Note that the gauge symmetry breaking scale turns out to be ${\cal O}(10^{15})$ GeV with $M_{S}^2 >$0  \cite{Rehman:2018nsn}. So we left with $M_{S}^{2}<$0. Now  $am_{3/2}$ can be $am_{3/2}=$0 or $am_{3/2}<$0 or $am_{3/2}>$0.   
Now with $M_{S}^{2}<$0 and $am_{3/2}=$0, $M$ again turns out be ${\cal O}(10^{15})$ or less \cite{Rehman:2009yj}. In this way we left with two cases that is $M_{S}^{2}<$0 with  $am_{3/2}>$0 or $am_{3/2}<$0. 
The second case is also discussed in \cite{Rehman:2018nsn}. With these two combination, we plot for $\mathcal{N}$=1 and $\mathcal N$=2 with $m_{3/2}$= 1,10,100 and 1000 TeV. For $\mathcal N$=1 red curves show $am_{3/2}>$0 and blue curves represent $am_{3/2}<$0 and green curves represent $am_{3/2}>$0 and purple curves represent $am_{3/2}<$0 for $\mathcal N=$2. Plots in $r-M$ plane are shown in the upper panel, while the lower panel shows plots in $|M_{S}|-M$ plane. From the top panel, it is clearly seen that for $am_{3/2}<$0 that is blue curves and purple curves, $M\sim 10^{15}$ GeV which is not acceptable due to the proton decay problem, which requires $M\sim 10^{16}$ GeV. On the other hand for the case of $M_{S}^{2}<0$ and $am_{3/2}>$0 that is red and green curves, we have $M\sim 10^{16}$ GeV for $m_{3/2}$= 1,10,100 and 1000 TeV. We also notice the small values of tensor to scalar ratio $r$ is in the range of $10^{-15}$ to $10^{-8}$ (red and green curves).
The lower panel gives us ranges of the gauge symmetry breaking scale $M$ and Soft SUSY breaking mass of singlet $|M_{S}|$. 
We note that for $|M_{S}|<10^{7}$ GeV with $am_{3/2}<$0, $M$ is not sensitive to the values of $|M_{S}|$. As $|M_{S}|$ is keep increasing till $10^{10}$ GeV, $M$ also increases and approaches to a maximum value ($M\sim 4\times 10^{15}$ GeV for $m_{3/2}=$ 1000 TeV) and then starts going down again. On the other hand there is some cancellation takes place when $M_{S}^{2}<0$ and $am_{3/2}>$0.  In result of this cancellation, $M$ starts rising for $|M_{S}|\gtrsim 10^{6}$ GeV.
This shows that in SUSY hybrid inflation we can achieve $M$ of the order of ${10^{16}}$ GeV at the price of higher Soft SUSY breaking mass of singlet $|M_{S}|\gtrsim 10^{6}$ GeV for both $\mathcal N$=1 and $\mathcal N$=2 with $m_{3/2}$= 1,10,100 and 1000 TeV. These large values of soft masses correspond to the Split-SUSY scenario ~\cite{ArkaniHamed:2004fb,Ahmed:2019xon}.  



In the upper panel of Fig.~\ref{fig4}, we show our calculations in $|M_{S}|-|S_{0}|/m_{p}$ plane and lower panel presents plots in $|M_{S}|-\kappa$. Color coding is same as in Fig.~\ref{fig3}. From upper panel we see that the maximum variation in $|S_{0}|/m_{p}$ is from $5\times 10^{-4}$ to $0.005$ for the variation of $|M_{S}|$ between $10^{6}$ GeV and $10^{7}$ GeV. Other ranges of $|S_{0}|/m_{p}$ with different values of $m_{3/2}$ are also shown with maximum value of $0.005$. In the lower panel, we also note that the value of $\kappa$ varies between $10^{-7}$ to $10^{-2}$ with variation of $|M_{S}|$ values in the mass interval of $10^{6}$ GeV to $10^{9}$ GeV for both red and green curves. One can also note that in both Fig.~\ref{fig3} and Fig.~\ref{fig4} for $|M_{S}|\gtrsim 10^{10}$ GeV there is no difference in results between $am_{3/2}>0$ and $am_{3/2}<0$.

In addition to above calculations, we also like to estimate the values of the Hubble parameter during inflation $H$ in these models. In the slow-roll limit, the Hubble parameter depends on the value of $V$ and is given as

\begin{equation}
H^{2}=\frac{V}{3m_{p}^{2}},
\end{equation}
\begin{figure}[t!]
	\centering	
	\begin{subfigure} {0.7\textwidth}
		\includegraphics[width=1\linewidth]{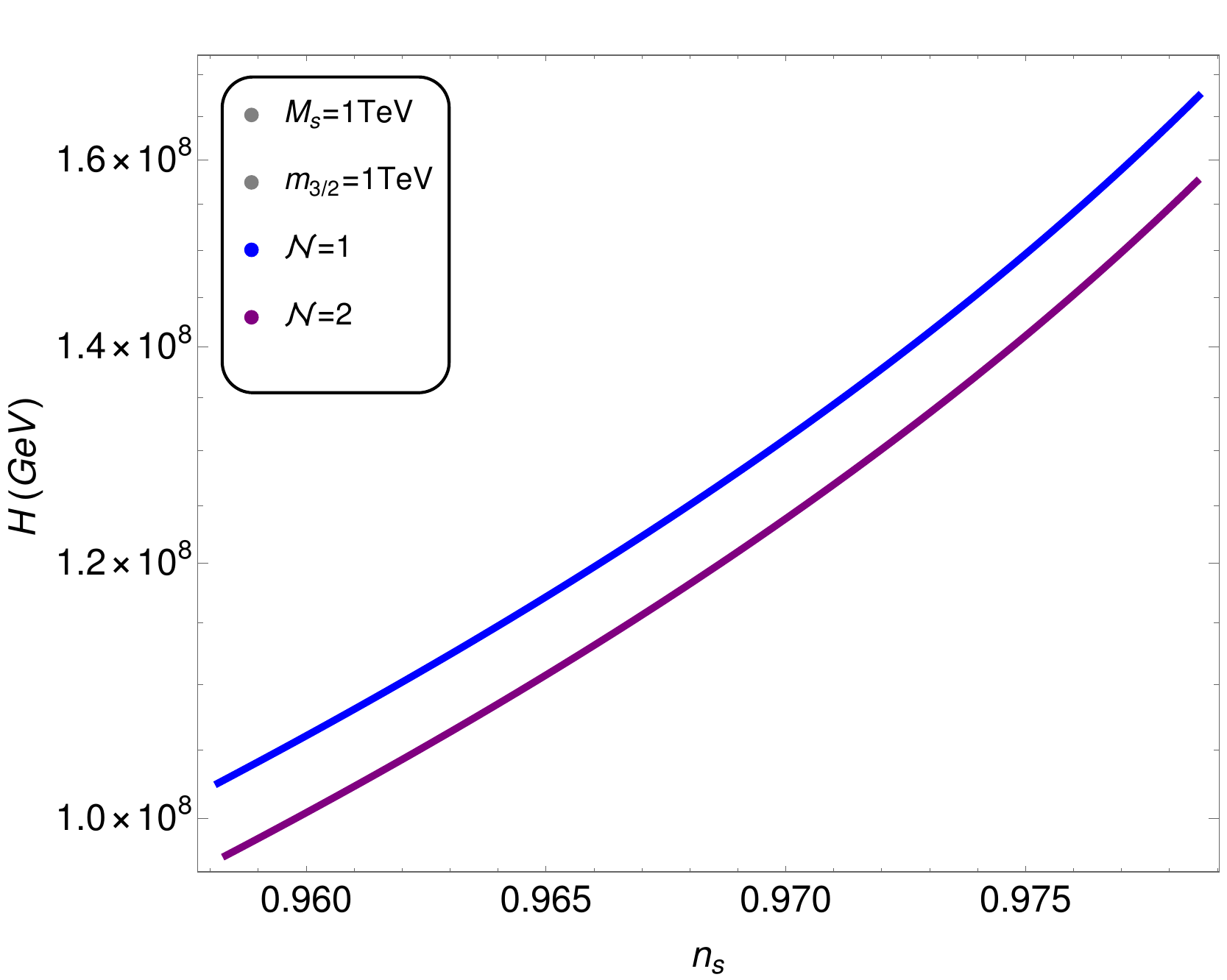}
	\end{subfigure}
	
	\caption{\small{Plot in $H-n_s$ plane with $\mathcal{N}=$1 (blue curve), $\mathcal{N}=$2 (purple curve) and $N_{0}=$50.}}
	\label{fig6}
\end{figure}
and we calculate it at the pivot scale. In Fig.~\ref{fig6} we show the values of the Hubble parameter in the ($n_{s}-H$) plane. Here we show our calculations for $\mathcal{N}$=1 and $\mathcal{N}$=2 as blue and purple lines respectively. We see that with the variation of 2$\sigma$ from $n_{s}$ central value, $H$ varies almost linearly from $9.9\times 10^{7}$ GeV to $1.68 \times 10^{8}$ GeV. In the context of modified swampland and TCC condition, it is hard to satisfy TCC conjectures in SUSY hybrid inflation with minimal K as TCC required very $H<1$ and very small $r\sim10^{-31}$ where the modified swampland conjecture, required c or $c^{\prime}$ of order 1 which can easily satisfied for $c^{\prime}$ case. However, to satisfy both conjectures simultaneously, we briefly discussed the SUSY hybrid inflation with non-minimal K, where we have large r and small r solutions.

\subsection{\large{\bf SUSY Hybrid Inflation with Non-Minimal K\"ahler Potential}}{\label{sec5}}
The non-minimal K\"ahler potential may be expanded as,
\bea
K &=& |S|^2 + |\Phi|^2 + |\overline{\Phi}|^2 + \frac{\kappa_S}{4}\frac{|S|^4}{m_P^2} + \frac{\kappa_\Phi}{4}\frac{|\Phi|^4}{m_P^2} +\frac{\kappa_{\overline{\Phi}}}{4}\frac{|\overline{\Phi}|^4}{m_P^2} \nonumber \\ 
 &+& \kappa_{S \Phi}\frac{|S|^2|\Phi|^2}{m_P^2}  +  \kappa_{S \overline{\Phi}}\frac{|S|^2|\overline{\Phi}|^2}{m_P^2} + \kappa_{\Phi \overline{\Phi}}\frac{|\Phi|^2|\overline{\Phi}|^2}{m_P^2} + \frac{\kappa_{SS}}{6}\frac{|S|^6}{m_P^4} + \cdots ,
\label{nonminkhlr}
\eea
Using Eqs.~\eqref{superpot}, \eqref{VF} and \eqref{nonminkhlr} along with the radiative correction and soft mass terms, we obtain the following inflationary potential,
\be
	V \simeq \kappa^{2}M^{4} \left( 1 - \kappa_S \left( \frac{M}{m_P} \right)^2 x^2 + \gamma_S\left( \frac{M}{m_{P}}\right)^{4}\frac{x^{4}}{2} + \frac{\kappa ^{2}\mathcal{N}}{8\pi ^{2}}F(x) \right. 
	\left. + a\left(\frac{m_{3/2}\,x}{\kappa\,M}\right) + \left( \frac{M_S\,x}{\kappa\,M}\right)^2\right),\label{scalarpot2}
\ee
where 
\be
\gamma _{S}=1-\frac{7\kappa _{S}}{2}+2\kappa _{S}^{2}-3\kappa _{SS} \label{gammas},
\ee
where $\gamma _{S}=1-\frac{7\kappa _{S}}{2}+2\kappa _{S}^{2}-3\kappa _{SS}$, and we have kept the terms up to $\mathcal{O}\; (\left(\vert S \vert / m_P \right)^4)$ from SUGRA corrections.
\begin{equation}
F(x)=\frac{1}{4}\left( \left( x^{4}+1\right) \ln \frac{\left( x^{4}-1\right)}{x^{4}}+2x^{2}\ln \frac{x^{2}+1}{x^{2}-1}+2\ln \frac{\kappa ^{2}M^{2}x^{2}}{Q^{2}}-3\right)
\end{equation}
and 
\begin{equation}
a = 2\left| 2-A\right| \cos [\arg S+\arg (2-A)].
\label{a}
\end{equation}
In the following discussion and numerical calculations, we mostly interested in large $r$ solution, modified swampland and trans-Plankian conjectures. So, we  consider the solution in which quadratic term positive and quartic is negative in potential. We further assume soft masses, $am_{3/2}= M_S= 1 TeV$, with $a$ negative sign.

\begin{figure}
	
	\begin{subfigure}{.5\textwidth}
		\centering	
		\includegraphics[width=1\linewidth]{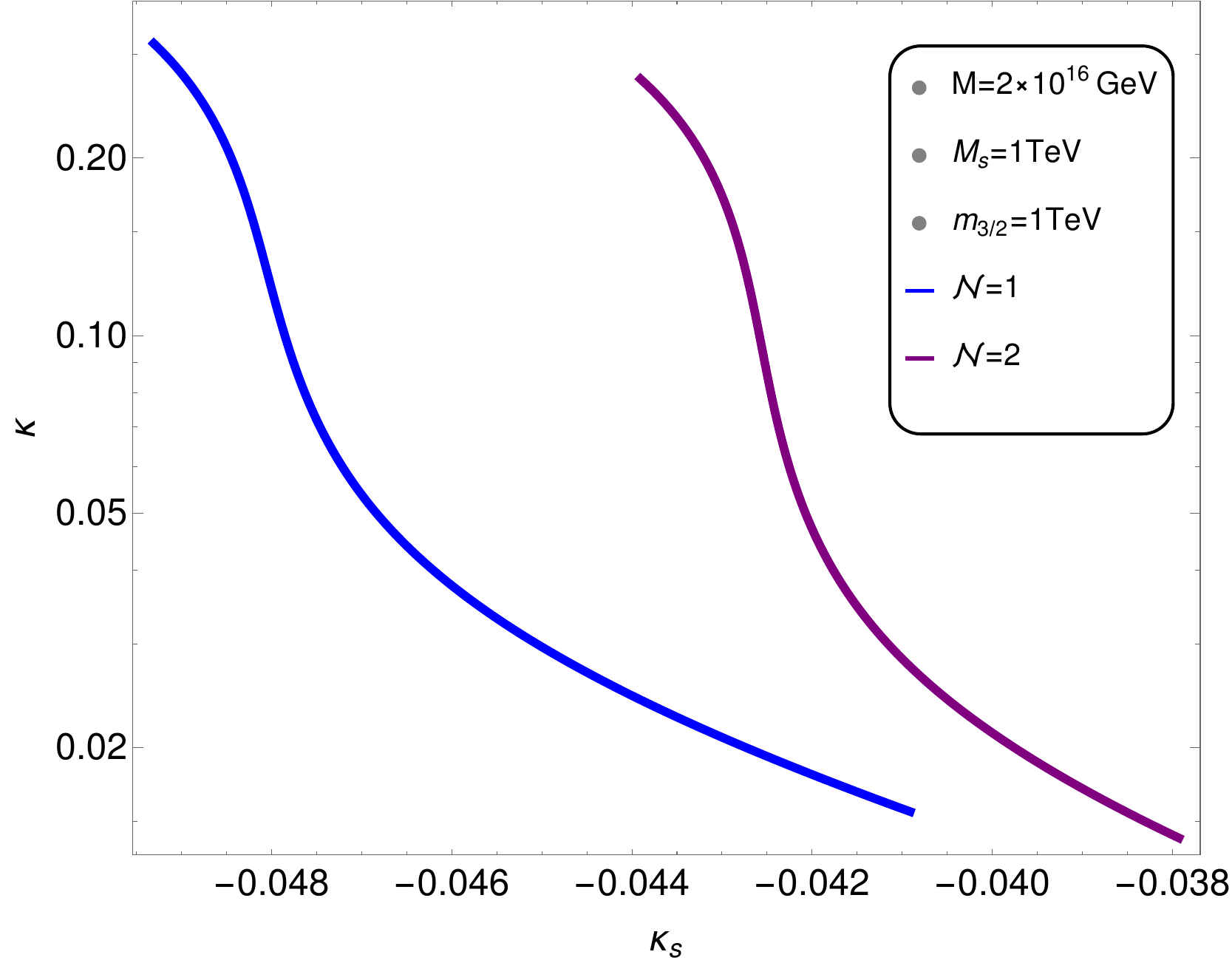}
		\label{4d2}
	\end{subfigure}
	\begin{subfigure}{.5\textwidth}
		\centering	
		\includegraphics[width=1\linewidth]{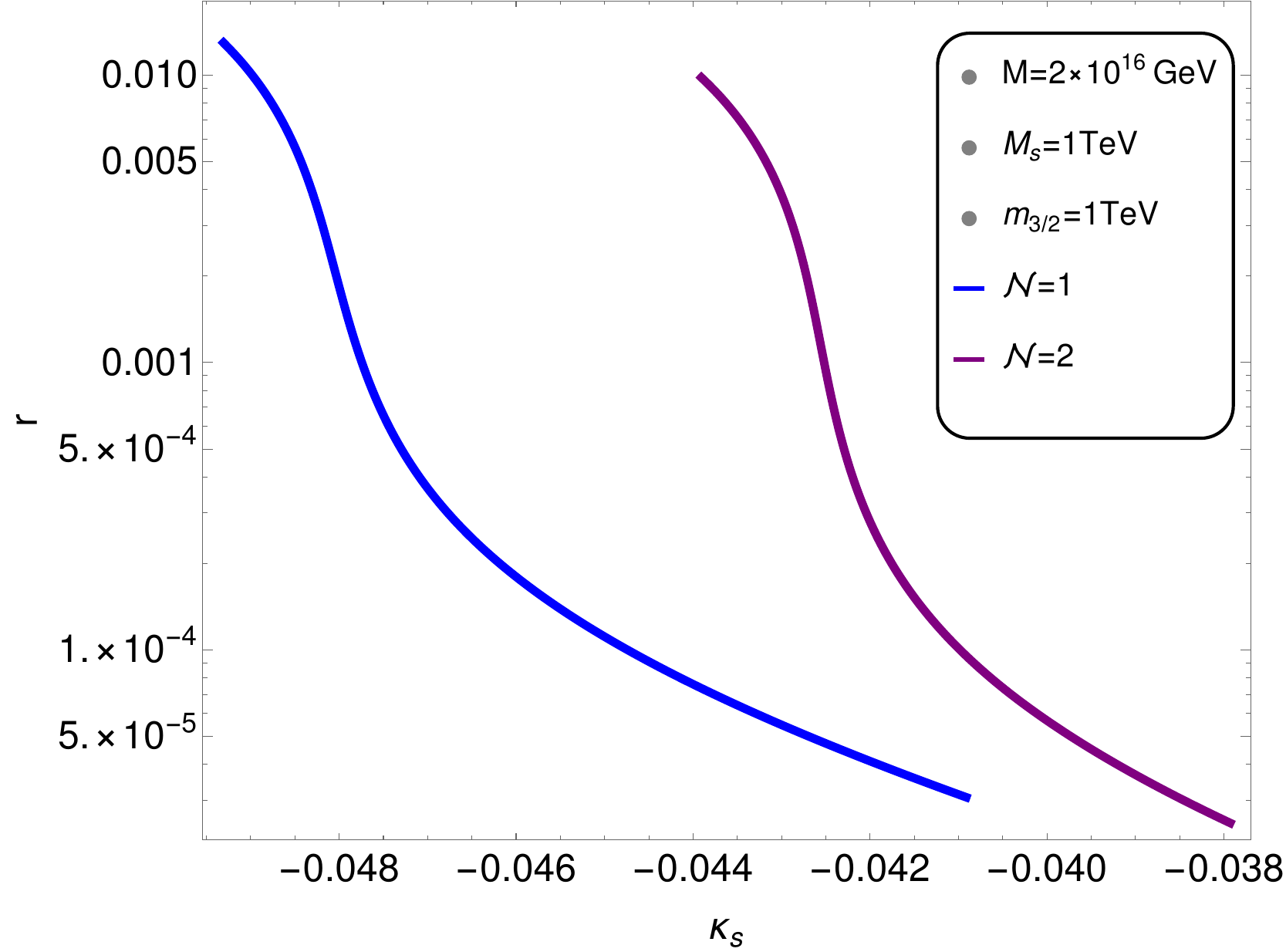}
		\label{4d2}
	\end{subfigure}\\
	\begin{subfigure}{.5\textwidth}
		\centering
		\includegraphics[width=1\linewidth]{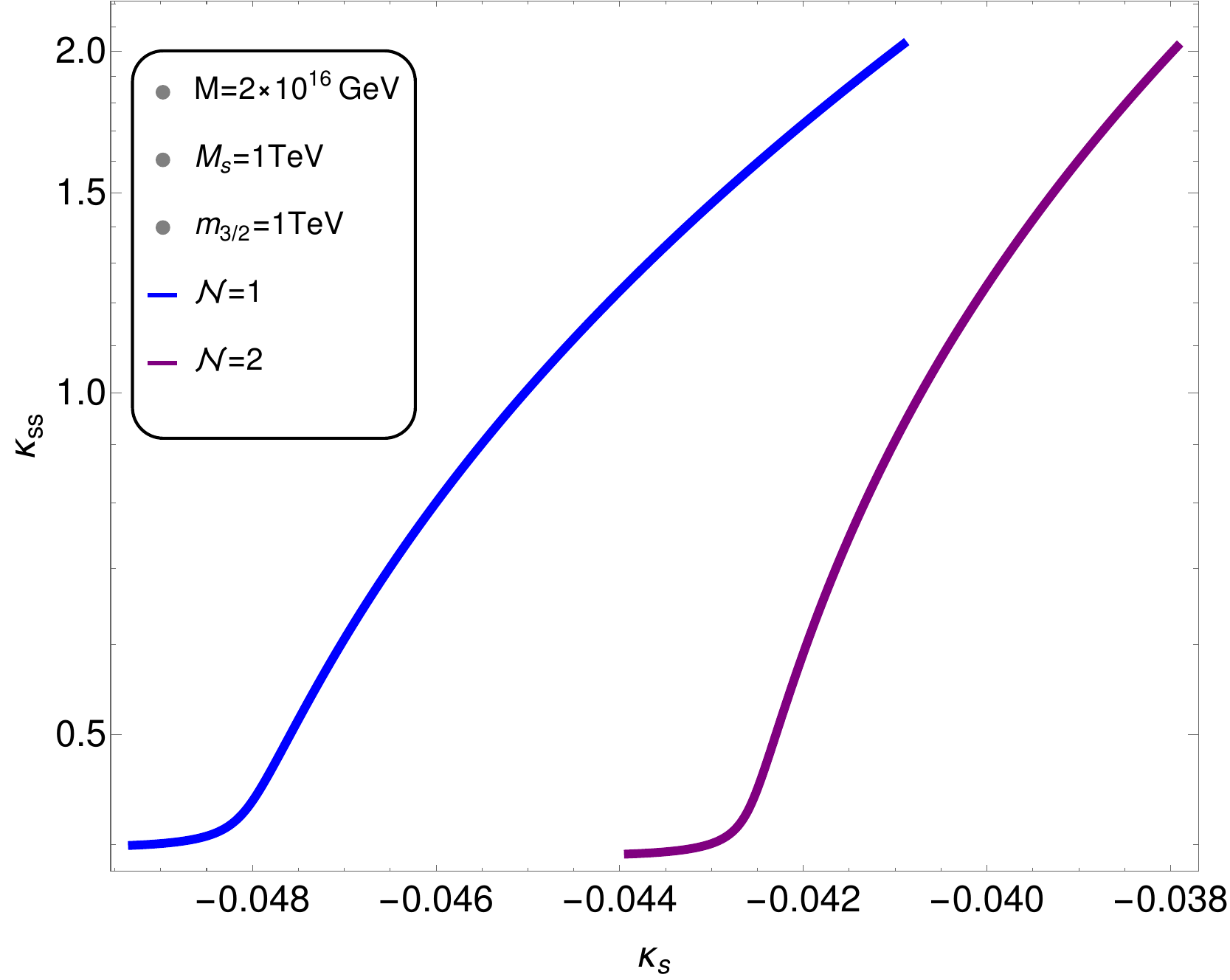}
		\label{4d2}
	\end{subfigure}	%
	\begin{subfigure}{.5\textwidth}
		\centering
		\includegraphics[width=1\linewidth]{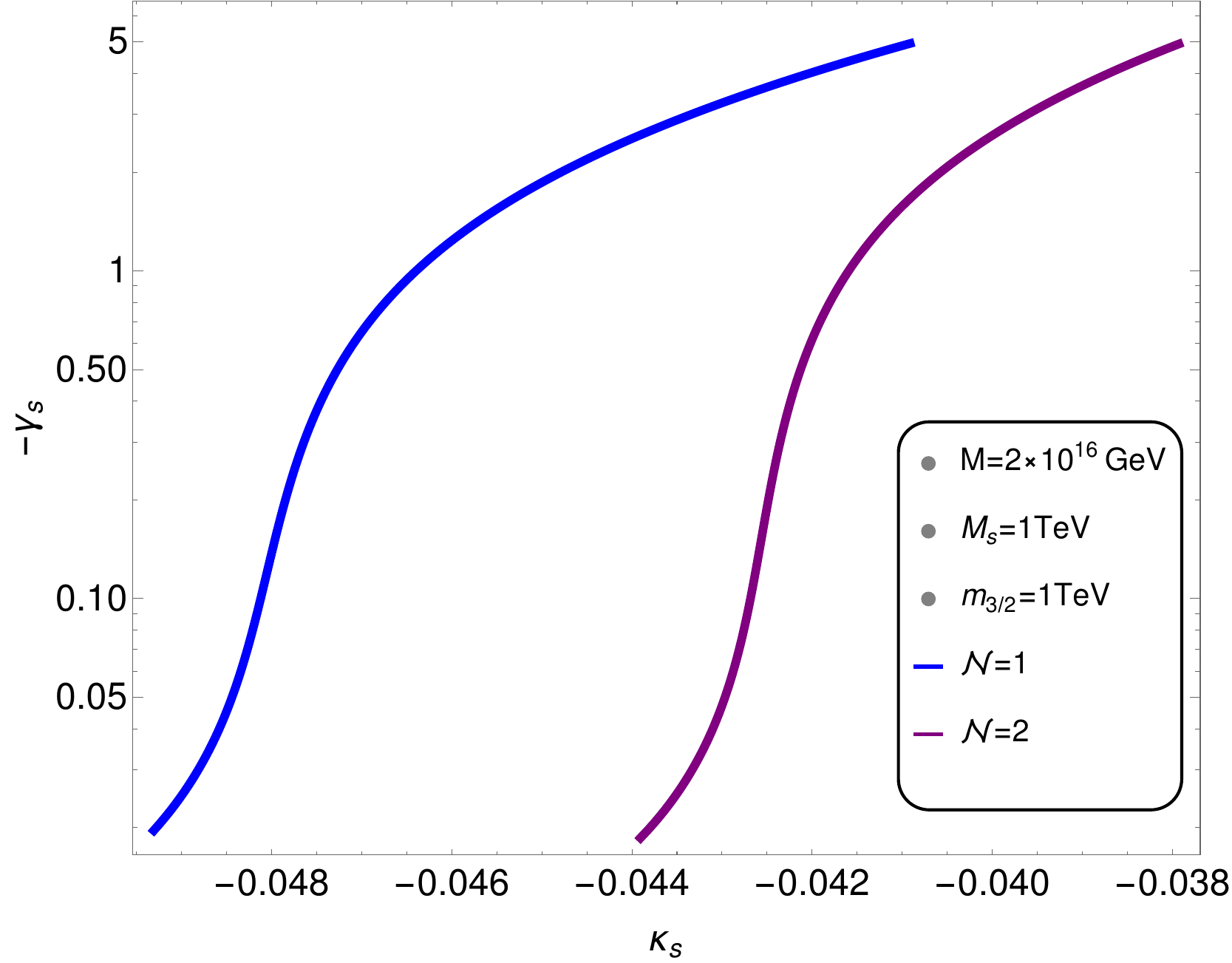}
		\label{4d1}
	\end{subfigure}
	
	\caption{{Plots for $\kappa$ (top left panel), tensor to scalar ratio $r$ (top right panel), $\kappa_{SS}$ (middle left panel) and $\gamma_{S}$ (middle right panel) with respect to the non-minimal coupling $\kappa_S$ for $N_0 = 50$, GUT symmetry breaking scale $M=2 \times 10^{16}$ GeV and $n_s = 0.9665$ (central value). Blue lines represent $\mathcal{N}=$ 1 while purple lines represents $\mathcal{N}=$2. }}
	\label{fig7}
\end{figure}

\begin{figure}[t!]
	
	\begin{subfigure}{.5\textwidth}
		\centering	
		\includegraphics[width=1\linewidth]{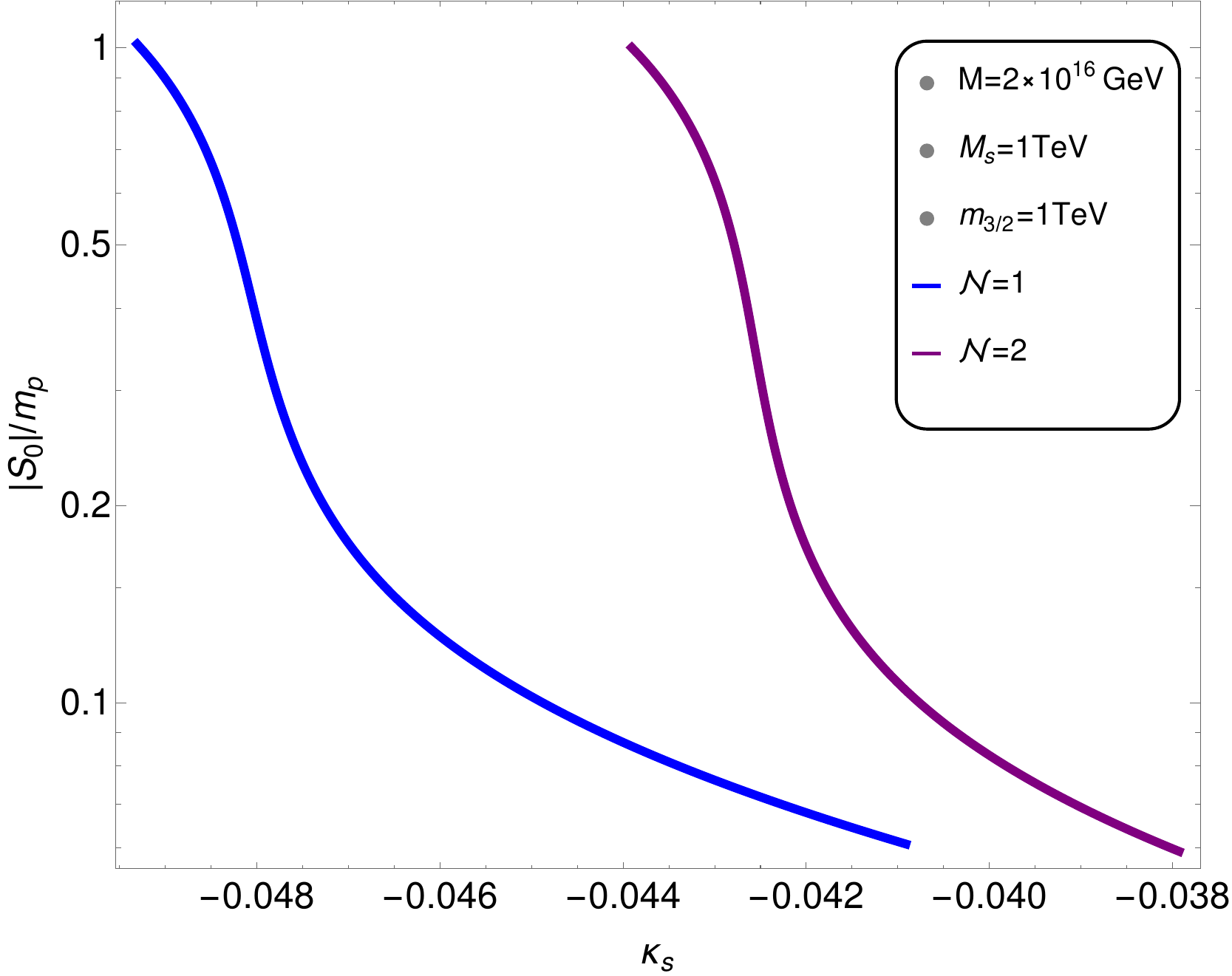}
	\end{subfigure}
	\begin{subfigure}{.5\textwidth}
		\centering	
		\includegraphics[width=1\linewidth]{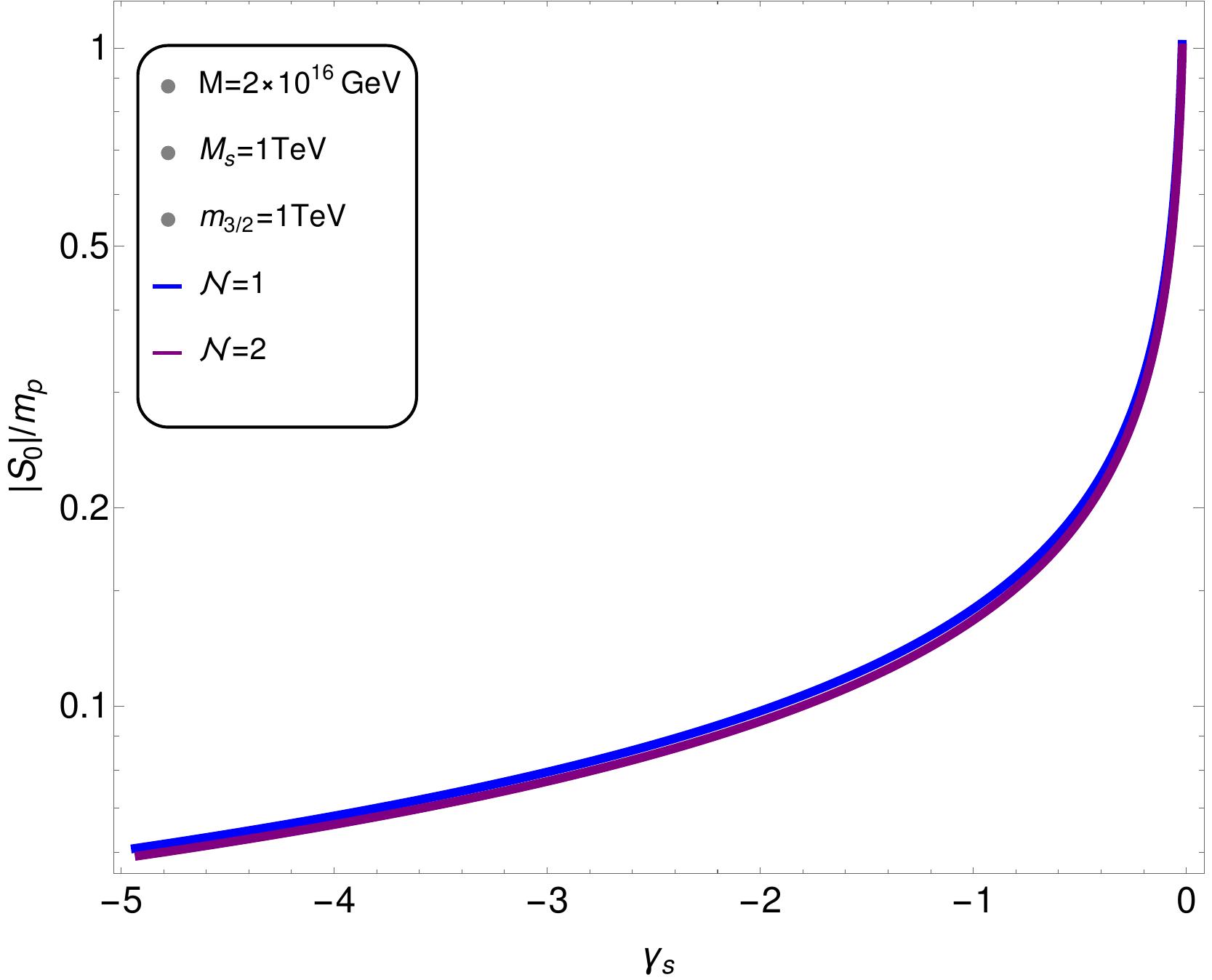}
	\end{subfigure}
	\caption{$\vert S_0 \vert / m_P$ versus non-minimal coupling $\kappa_S$ (left panel) and $\gamma_S$ (right panel). Color coding is same as in Fig.~\ref{fig7}.}
	\label{fig8}
\end{figure}

\subsubsection{Large r Solutions}
 
 .
 For our calculations shown in Fig.~\ref{fig7}, we assume soft masses, $am_{3/2}= M_S= 1 $ TeV, with $a$ negative sign. We display our calculations for $\mathcal N=$1 (blues curves) and $\mathcal N=$2 (purple curves). We set $n_{s}=$0.9665 (central value) and number of e-folding $N_{0}=$50.
In these figures all the solutions shown have sub-Planckian field values. From the left panel of Fig.~\ref{fig7}, we note that $\kappa$ have large values $[0.015,0.25]$ for $\kappa_{S}$ ranges $[-0.05,-0.041]$ (blue curve) and  we also note that $\kappa$ have range $[0.01,0.25]$ with $\kappa_{S}$ in the range $[-0.044,-0.038]$ (purple curve). The large values of $\kappa$ also implies large values of $r$. This can be understood from the relation $r \simeq \left( \frac{2  \kappa^2}{3 \pi^2 A_s (k_0)} \right) \left( \frac{M}{m_P} \right)^4$ given in \cite{Rehman:2018nsn}. This relation implies that since we have $M=2\times 10^{16}$ GeV, for large values of $\kappa$, we will also have large values of $r$. For the similar ranges of $\kappa_{S}$ as stated before, from the right panel we see that the maximum value of $r$ is about $0.01$ for both $\mathcal{N}=$1 and $\mathcal{N}=$2 and it can be as small as $10^{-5}$.  The large tensor modes predicted by the model are potentially measurable by forthcoming CMB experiments such as, LiteBIRD \cite{LiteBIRD:2022cnt}, Simons Observatory \cite{SimonsObservatory:2018koc}, PRISM \cite{Andre:2013afa}, CMB-S4 \cite{CMB-S4:2020lpa}, CMB-HD \cite{Sehgal:2019ewc}, CORE \cite{Finelli:2016cyd} and PIXIE \cite{Kogut:2011xw}.

In the lower left and right panels we plot $\kappa_{SS}$ and $\gamma_{S}$ respectively as a function of $\kappa_{S}$. Here we want to comment that with $\kappa_{S}<$0, we do not have solutions for $\gamma_{S}>$0 with $n_{s}=$0.9665.
In addition to it, in this case, we have solutions only for $\kappa_{SS}>$0 as for $\kappa_{SS}<$0, fields become trans-Planckian. From the plot in $\kappa_{S}-\kappa_{SS}$ plane we see that for $\mathcal{N}=$1 and $\mathcal{N}=$2, we have almost same range for $\kappa_{SS}$ $[0.4,2]$ for two different ranges of $\kappa_{S}$ that is  from $-0.049$ to $-0.041$ (blue curve) and from $-0.044$ to $-0.038$ (purple curve). 
On the other hand, in $\kappa_{S}-\gamma_{S}$ plot, we see that for similar ranges of $\kappa_{S}$, $\gamma_{S}$ varies -5 to -0.02 for $\mathcal{N}=$1 and $\mathcal{N}=$2. Finally, Fig.~\ref{fig8} shows the variation of $|S_{0}|/m_{p}$ with respect to $\kappa_{S}$ (left panel) and $\gamma_{S}$ (right panel). From the right panel, we see that when $\kappa_{S}$ varies between $-0.049$ to $-0.041$, $|S_{0}|/m_{p}$ changes between $0.06$ to $1$. We see from the right panel that $|S_{0}|/m_{p}$ changes between $0.06$ to $1$ when we change $\gamma_{S}$ from $-5$ to $0$.

\subsection{Swampland Conjectures}
\begin{figure}
	\begin{subfigure}{.5\textwidth}
		\centering	
		\includegraphics[width=1\linewidth]{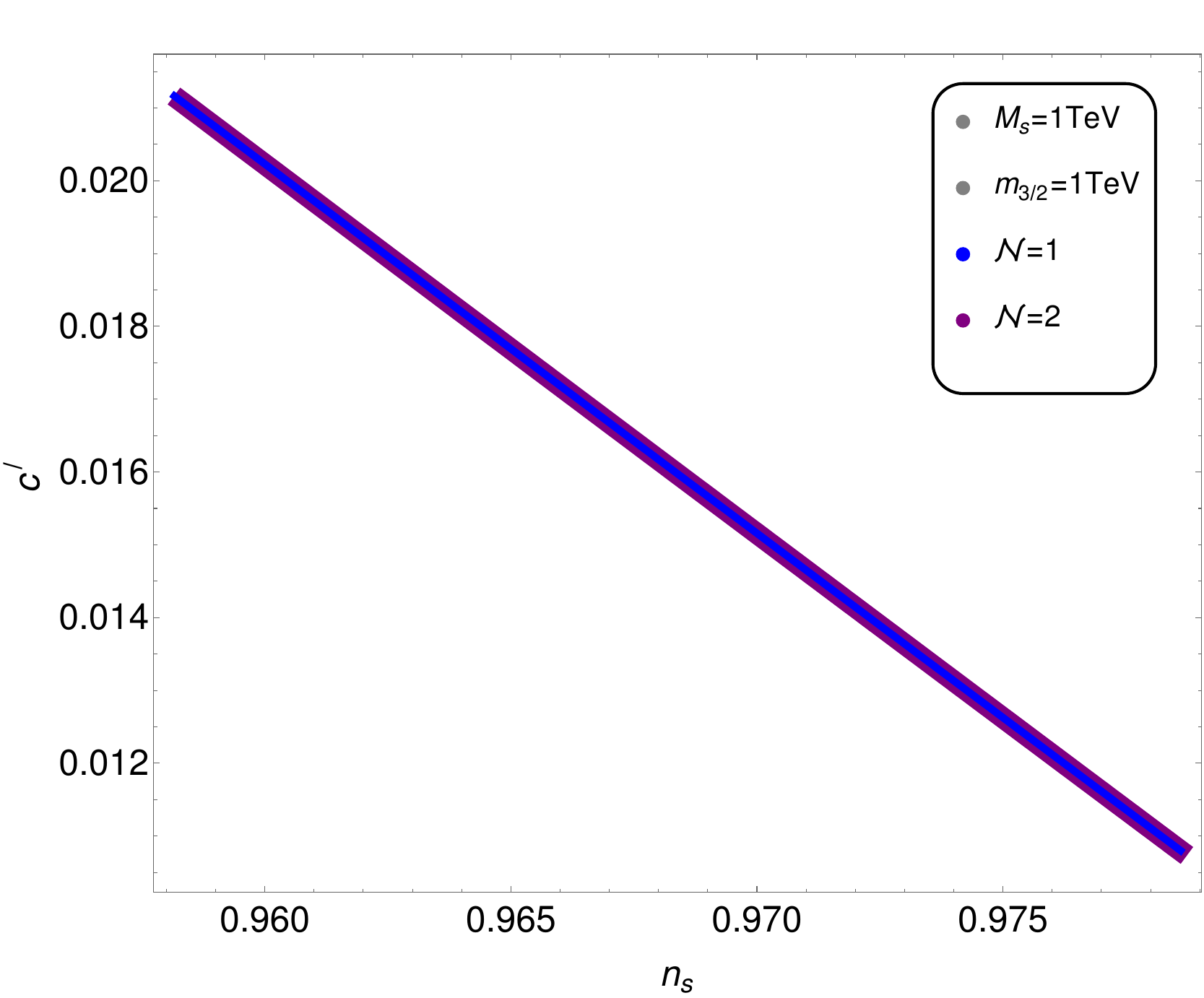}
	\end{subfigure}%
	\begin{subfigure}{.5\textwidth}
		\centering	
		\includegraphics[width=1\linewidth]{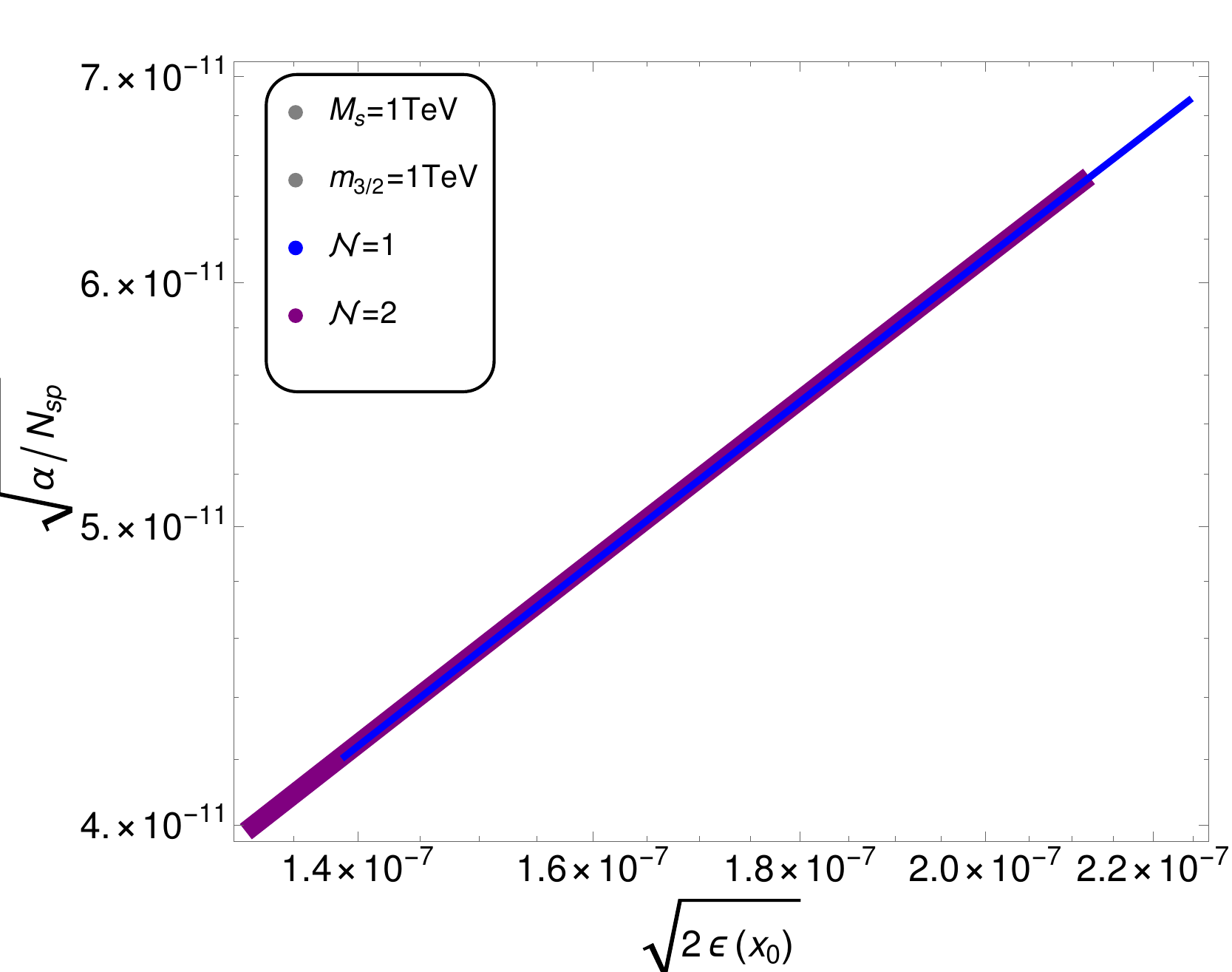}
	\end{subfigure}
	\caption{\small{Figure shows the plots for  $n_{s}$ v.s $-c^{\prime}$ (left panel) and $\sqrt{\alpha/N_{sp}}$ v.s $\sqrt{2\epsilon(x_{0})}$ for minimal case. Color coding is same as in Fig.~\ref{fig5}.}}
	\label{fig5}
\end{figure}

\begin{figure}[t!]
  	\centering	
  	\begin{subfigure} {0.5\textwidth}
  		\includegraphics[width=1\linewidth]{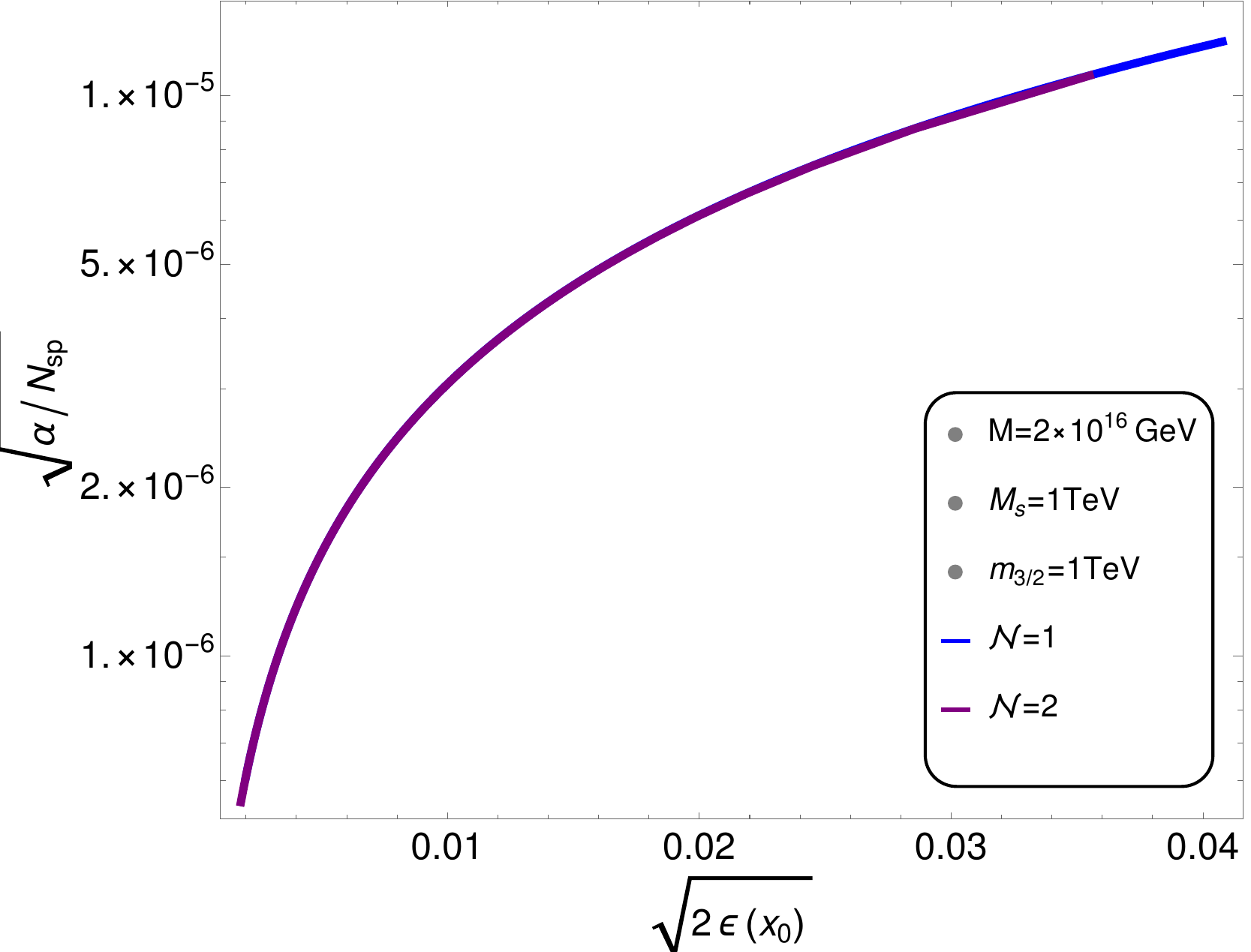}
  	\end{subfigure}
  	\caption{\small{Figure shows the plot for  $\sqrt{\alpha/N_{sp}}$ v.s $\sqrt{2\epsilon(x_{0})}$ for $\mathcal{N}$=1 (blue curve) and  $\mathcal{N}$=2 (purple curve) for non-minimal case. }}
  	\label{fig9}
  \end{figure}

{\label{sectcc}}
In addition to the basic inflationary predictions we also discuss the bounds on parameter space coming from swampland conjecture as shown in Fig.~\ref{fig5} where in the left panel we show $n_{s}$ vs $-c^{\prime}$  plane and in the right panel we display plot in $\sqrt{2\epsilon}$ vs $\sqrt{\alpha/N_{sp}}$ plane. The swampland conjecture put the lower bound on tensor to scaler ratio $r$ as discussed equation in ~\ref{D-dSC}. The detail of this bound can be seen below \footnote{{$2\epsilon\gtrsim \alpha$ $\Rightarrow$ $\frac{r}{8}\gtrsim \alpha$}}. 
Recall that the $-c^{\prime}$ which is actually our $\eta$. We calculate $-c^{\prime}$ at pivot scale. So the range of $-c^{\prime}$ at pivot scale is $(-0.023,-0.01 )$. Also the inflation ends at $\eta(x_{e})=-1$. So $-c^{\prime}$ have range from pivot scale to end of inflation which is $(-0.023,-1)$ for different $\mathcal{N}$. From Fig.~\ref{fig5} (left panel), it is clear that the values of $-c^{\prime}$ varies from -0.021 to -0.011 which is not an $O(1)$ number. On the other hand as we discussed before, according to Dvali et al.~\cite{Dvali:2018jhn}, the parameter $\alpha$ is related to Vafa et al.~\cite{Ooguri:2018wrx} parameter $c$ as $\sqrt{\alpha}=c$ but as compare to $c$ which is $O(1)$ parameter, $\sqrt{\alpha}$ can be any number $< 1$. Recall that as we discussed in \ref{sec3} that $\sqrt{\alpha}=c=\sqrt{2\epsilon(x_{0})}$.
Now one should note that according to equation~\ref{V-RdS}, any one of the two conditions can be satisfied to remain consistent with modified swampland conjecture. So following equation~\ref{V-RdS} and \ref{D-dSC}, we present our calculations in the right panel of Fig.~\ref{fig5}. Here we see that $\sqrt{{\alpha}/N_{sp}}$ and $\sqrt{2\epsilon(x_{0})}$ are of order $10^{-11}$ and $10^{-7}$ respectively and thus our results are consistent with the modified swampland conjecture for both $\mathcal{N}=$1 and $\mathcal{N}=$2 for minimal SUSY hybrid inflation. In Fig.~\ref{fig9}, we show our calculations for swampland conjecture employing non-minimal SUSY hybrid inflation model with $N_{0}=50$. We see that with the variation of $\sqrt{2\epsilon}$ between 0.002 to 0.04, $\sqrt{\alpha/N_{sp}}$ changes from ${\cal O}(10^{-6})$ to ${\cal O}(10^{-5})$ which is consistent with swampland conjectures.


\subsection{TCC Conjecture} 
The TCC conjecture yields an approximate upper bound of around $10^9$ GeV on the energy scale during inflation, or equivalently, the upper bound on the Hubble scale $H \leq 1$ GeV during the (slow-roll) inflation. To discuss the TCC conjecture briefly, we consider the potential \ref{gammas} up to quartic terms by making delicate cancellations between $\kappa_{S}$ and $M_{S}$ terms, so the remaining potential behaves like the inflection-point inflation (IPI) scenario. Here we consider the leading order terms for our analytics; otherwise, higher-order terms distort the inflection-point inflation. We need to do more fine-tuning in parameters of higher-order terms to maintain these results, which we derive below. By consider the amplitude of primordial perturbation $A_{s}$ with tensor-to-scalar $r$ , we can find $\kappa$ in terms of $r$ and $A_{s}$. Similarly, the delicate cancellations between $\kappa_{S}$ and $M_{S}$ terms give us a relation between $\kappa_{S}$ and $M_{S}$ which reads as,
 \bea
 \kappa=\sqrt{\frac{3\pi^2 A_{s} r}{2\left(\frac{M}{m_{p}}\right)^4}} , \qquad  \kappa_{S}=\left(\dfrac{M_{S} m_{p}}{\kappa M^2}\right)^2.
 \eea \label{wa1}
 Where $r$ is the value of tensor-to-scalar ratio at pivot scale. The scalar spectral index $n_s$ defined in eq \ref{ns} for small $r$ solution we can write $n_{s}$ and $r$ as follow, 
 \bea
 n_{s}\approx 1+2\eta \Rightarrow \gamma_{s}=\frac{n_{s}-1}{6 x_{0}^2}\left(\frac{m_{p}}{M}\right)^2, \quad r\approx16 \epsilon \Rightarrow 4\left(\frac{m_{p}}{M}\right)^2\left(\frac{am_{3/2}}{\kappa M}-\frac{2\gamma_{s}M^4x_{0}^3}{m_{p}^4}\right)^2.
 \eea \label{wa2}
 Using above two equation, we can get a relation between $r$ and $n_{s}$ at pivot scale,
 \bea
r= 4\left(\frac{m_{p}}{M}\right)^2\left(\frac{am_{3/2}}{\kappa M}-\frac{\left(n_{s}-1\right)x_{0}}{3}\left(\frac{M}{m_{p}}\right)^2\right)^2.
 \eea \label{wa3}
 The benchmark points for TCC conjectures are shown in table \ref{table1}. Point 1, 2 and 3 shows the solutions consistency with TCC bounds on the energy density $E_{inf}$ and scale of inflation H. \footnote{We note that in recent years, several works have explored the context of Swampland conjectures and their implications on inflation models \cite{Blumenhagen:2017cxt,Chiang:2018lqx, Scalisi:2018eaz, Haque:2019prw, Berera:2019zdd, Li:2019ipk, NooriGashti:2021nox, Gashti:2022pvu, Osses:2021snt, Kamali:2021ugx, Sabir:2019wel}.}
 
 \begin{table}\hspace{1.0cm}
	\centering
	\begin{tabular}{|c|ccc|}
		\hline
		\hline
		& Point 1 & Point 2 & Point 3 \\

		\hline

		$M (GeV)$        & $2\times10^{16}$  & $2\times10^{16}$   & $2\times10^{16}$ \\
		\hline
		$M_{S}(GeV)$        & 0.1  & 0.1   & 0.1   \\
		\hline
		$\kappa_{S}$       & 0.472509 &  0.0472509  & $0.00472509$   \\
		\hline
		$\kappa$        & $8.86\times 10^{-16}$ &  $8.86\times 10^{-16}$   & $8.89\times 10^{-15}$ \\
		\hline
		$\gamma_{s}$ & -85.2687 & -85.2687   & -85.2687 \\
		\hline
		$N_{0}$  &45.363 &  44.212  & 44.2117   \\
		\hline
		
		$am_{3/2}(GeV)$    & 0.0000137383  & 0.0000434442           & 0.000137383  \\
		\hline
		$E_{inf}(GeV)$   & $5.95\times 10^8$  & $1.06\times10^9$ & $1.88\times 10^9$  \\
		\hline
		$H(GeV)$ &0.0484924 &0.153346 & 0.484924 \\
		\hline
		$n_{s}$       &0.9655 &0.9655& 0.9655 \\
		\hline
		$r$  &$9.63\times 10^{-32}$ &$1\times10^{-30}$ &$1\times10^{-29}$ \\
		
		\hline
		
		\hline
	\end{tabular}
	\caption{Table shows the bench mark points of the TCC conjectures. The inflaton field at pivot scale $x_{0}$ is close to $x_{e}$ which is close to 1.
	}
	\label{table1}
\end{table}

\section{\large \bf {Summary}}{\label{sec6}}
In this study, we revisit supersymmetric (SUSY) hybrid inflation in light of CMB experiments and swampland conjecture. We first show that if one adds radiative and soft mass corrections to the scalar potential along with SUGRA corrections, supersymmetric hybrid inflation is still consistent with Planck 2018 data despite the impression that it does not. Usually, in SUSY hybrid inflation with minimal K\"ahler potential, the gauge symmetry breaking scale $M$ turns out to be ${\cal O}(10^{15})$ GeV, which causes proton decay rate problem. In this article, we present a new parameter space where the proton decay rate problem can be avoided by achieving $M$ of the order of $10^{16}$ GeV with $M_{S}^{2}<$0 and $am_{3/2}>$0. In this scenario, one requires a soft SUSY breaking scale $|M_{S}| \gtrsim 10^{6}$ GeV. In addition to it, the tensor to scalar ratio $r$ is in the range $10^{-16}$ to $10^{-6}$, which is quite small. The modified swampland conjecture holds, whereas TCC conjecture is hard to satisfy. For this reason, we also consider non-minimal K\"ahler potential. We fixed spectral index $n_{S}=$0.9665 (central value) of Planck 2018 data and $M=2\times 10^{16}$ GeV and present our calculations. We show that with $M_{S}=$ 1 TeV, $m_{3/2}=$ 1 TeV, $\kappa_{S}<0$ for $\cal{N}=$1 and $\cal{N}=$2, $r$ ranges from $10^{-5}$ to $0.01$. We also present a parametric space and benchmark points table for SUSY Hybrid Inflation with non-minimal K\"ahler potential to show consistency with modified swampland and TCC conjecture.

\section{Acknowledgement}
We would like to thanks Gia Dvali for fruitful discussion. We would also like to thanks to Qaisar Shafi and Mansoor Ur Rehman for their valuable discussion at the beginning of the project. Their insightful discussions were particularly beneficial during the calculations.


\begin{thebibliography}{99}

\bibitem{Aad:2012tfa}
  G.~Aad {\it et al.}  [ATLAS Collaboration]
  Phys.\ Lett.\ B {\bf 716}, 1 (2012)
  [arXiv:1207.7214 [hep-ex]].

\bibitem{CMS} 
  S.~Chatrchyan {\it et al.}  [CMS Collaboration],
  Phys.\ Lett.\ B {\bf 716}, 30 (2012)
  [arXiv:1207.7235 [hep-ex]].


\bibitem{Aaboud:2017vwy} 
  M.~Aaboud {\it et al.} [ATLAS Collaboration],
  Phys.\ Rev.\ D {\bf 97}, no. 11, 112001 (2018)
  [arXiv:1712.02332 [hep-ex]].


\bibitem{Sirunyan:2017cwe}
  A.~M.~Sirunyan {\it et al.} [CMS Collaboration],
  Phys.\ Rev.\ D {\bf 96} (2017) no.3,  032003
  [arXiv:1704.07781 [hep-ex]].


\bibitem{Sirunyan:2017kqq} 
  A.~M.~Sirunyan {\it et al.} [CMS Collaboration],
  Eur.\ Phys.\ J.\ C {\bf 77}, no. 10, 710 (2017)
E.Kearns, (2015)
\url{http://atlas.web.cern.ch/Atlas/GROUPS/PHYSICS/HIGGS/higgs-xsec/cross.pdf}
  [arXiv:1705.04650 [hep-ex]].
  




\bibitem{Aprile:2018dbl} 
  E.~Aprile {\it et al.} [XENON Collaboration],
  Phys.\ Rev.\ Lett.\  {\bf 121}, no. 11, 111302 (2018)
  [arXiv:1805.12562 [astro-ph.CO]].  
  
\bibitem{Hinshaw:2012aka}
  G.~Hinshaw {\it et al.} [WMAP Collaboration],
  Astrophys.\ J.\ Suppl.\  {\bf 208}, 19 (2013)
  doi:10.1088/0067-0049/208/2/19
  [arXiv:1212.5226 [astro-ph.CO]].
Schmitz:2019uti

\bibitem{Akrami:2018odb} 
  Y.~Akrami {\it et al.} [Planck Collaboration],
  arXiv:1807.06211 [astro-ph.CO].




\bibitem{Kearns}
E.Kearns, (2015)
\url{http://atlas.web.cern.ch/Atlas/GROUPS/PHYSICS/HIGGS/higgs-xsec/cross.pdf}


\bibitem{Dvali:1994ms}
  G.~R. Dvali, Q.~Shafi, and R.~K. Schaefer,
  Phys. Rev. Lett. {\bf 73}, 1886 (1994), arXiv:hep-ph/9406319.
  
  \bibitem{Copeland:1994vg}
  E.~J. Copeland, A.~R. Liddle, D.~H. Lyth, E.~D. Stewart, and D.~Wands,
  Phys. Rev. {\bf D49}, 6410 (1994), arXiv:astro-ph/9401011.
  
  \bibitem{Linde:1993cn}
  A.~D. Linde,
  Phys. Rev. {\bf D49}, 748 (1994), arXiv:astro-ph/9307002.

  \bibitem{Dvali:1997uq}
  G.~R. Dvali, G.~Lazarides, and Q.~Shafi,
  Phys. Lett. {\bf B424}, 259 (1998), arXiv:hep-ph/9710314.

\bibitem{Chamseddine:1982jx} 
  A.~H.~Chamseddine, R.~L.~Arnowitt and P.~Nath,
  ``Locally Supersymmetric Grand Unification,''
  Phys.\ Rev.\ Lett.\  {\bf 49}, 970 (1982);
  R.~Barbieri, S.~Ferrara and C.~A.~Savoy,
  ``Gauge Models with Spontaneously Broken Local Supersymmetry,''
  Phys.\ Lett.\  {\bf 119B}, 343 (1982);
  L.~J.~Hall, J.~D.~Lykken and S.~Weinberg,
  ``Supergravity as the Messenger of Supersymmetry Breaking,''
  Phys.\ Rev.\ D {\bf 27}, 2359 (1983);
  E.~Cremmer, P.~Fayet and L.~Girardello,
  ``Gravity Induced Supersymmetry Breaking and Low-Energy Mass Spectrum,''
  Phys.\ Lett.\  {\bf 122B}, 41 (1983).
  N.~Ohta,
  ``Grand Unified Theories Based On Local Supersymmetry,''
  Prog.\ Theor.\ Phys.\  {\bf 70}, 542 (1983).



\bibitem{Jeannerot:2000sv} 
  R.~Jeannerot, S.~Khalil, G.~Lazarides and Q.~Shafi,
  JHEP {\bf 0010}, 012 (2000)
  [hep-ph/0002151].Congradulations

  \bibitem{Ade:2015lrj} 
  P.~A.~R.~Ade {\it et al.} [Planck Collaboration],
  Astron.\ Astrophys.\  {\bf 594}, A20 (2016)
  [arXiv:1502.02114 [astro-ph.CO]].



\bibitem{rehman}
M.U.~Rehman, Q.~Shafi, and J.R.~Wickman, Phys. Lett. 
B {\bf 683}, 191 (2010);
C.~Pallis and Q.~Shafi, Phys. Lett. B {\bf 725}, 327 
(2013);
W.~Buchm\"{u}ller, V.~Domcke, K.~Kamada, and K.~Schmitz, 
J. Cosmol. Astropart. Phys. {\bf 07}, 054 (2014).

\bibitem{gravitywaves}
M.U.~Rehman, Q.~Shafi, and J.R.~Wickman, Phys. Rev. D 
{\bf 83}, 067304 (2011); 
M.~Civiletti, C.~Pallis, and Q.~Shafi, Phys. Lett. B 
{\bf 733}, 276 (2014).

\bibitem{bastero}
M.~Bastero-Gil, S.F.~King, and Q.~Shafi, Phys. Lett. 
B {\bf 651}, 345 (2007).



\bibitem{LiteBIRD:2022cnt}
E.~Allys \textit{et al.} [LiteBIRD],
PTEP \textbf{2023} (2023) no.4, 042F01
doi:10.1093/ptep/ptac150
[arXiv:2202.02773 [astro-ph.IM]].

\bibitem{Agrawal:2018own} 
  P.~Agrawal, G.~Obied, P.~J.~Steinhardt and C.~Vafa,
  Phys.\ Lett.\ B {\bf 784}, 271 (2018)
  [arXiv:1806.09718 [hep-th]].



\bibitem{Obied:2018sgi} 
  G.~Obied, H.~Ooguri, L.~Spodyneiko and C.~Vafa,
  arXiv:1806.08362 [hep-th].







\bibitem{Ooguri:2018wrx} 
  H.~Ooguri, E.~Palti, G.~Shiu and C.~Vafa,
  Phys.\ Lett.\ B {\bf 788}, 180 (2019)
  [arXiv:1810.05506 [hep-th]].


\bibitem{Andriot:2018mav} 
D.~Andriot and C.~Roupec,
Fortsch.\ Phys.\  {\bf 67}, no. 1-2, 1800105 (2019)
[arXiv:1811.08889 [hep-th]]
\bibitem{Guth:1980zm} 
  A.~H.~Guth,
  Phys.\ Rev.\ D {\bf 23}, 347 (1981)
  [Adv.\ Ser.\ Astrophys.\ Cosmol.\  {\bf 3}, 139 (1987)].
  A.~D.~Linde,
  Phys.\ Lett.\  {\bf 108B}, 389 (1982)
  [Adv.\ Ser.\ Astrophys.\ Cosmol.\  {\bf 3}, 149 (1987)].
  A.~Albrecht and P.~J.~Steinhardt,
  Phys.\ Rev.\ Lett.\  {\bf 48}, 1Scalisi:2018eaz220 (1982)
  [Adv.\ Ser.\ Astrophys.\ Cosmol.\  {\bf 3}, 158 (1987)].
  A.~A.~Starobinsky,
  Phys.\ Lett.\ B {\bf 91}, 99 (1980)
  [Phys.\ Lett.\  {\bf 91B}, 99 (1980)]
  [Adv.\ Ser.\ Astrophys.\ Cosmol.\  {\bf 3}, 130 (1987)].
  K.~Sato,
  Mon.\ Not.\ Roy.\ Astron.\ Soc.\  {\bf 195}, 467 (1981).
\bibitem{Ahmed:2018jlv} 
W.~Ahmed and A.~Karozas,
Phys.\ Rev.\ D {\bf 98}, no. 2, 023538 (2018)
[arXiv:1804.04822 [hep-ph]].

\bibitem{Schmitz:2019uti} 
K.~Schmitz,
arXiv:1910.08837 [hep-ph].
  E.~Palti,
  arXiv:1903.06239 [hep-th].
  T.~D.~Brennan, F.~Carta and C.~Vafa,
  PoS TASI {\bf 2017}, 015 (2017)
  [arXiv:1711.00864 [hep-th]].
Z.~Yi and Y.~Gong,
arXiv:1811.01625 [gr-qc].

M.~Sabir, W.~Ahmed, Y.~Gong, T.~Li and J.~Lin,
arXiv:1908.05201 [hep-ph].
M.~Sabir, W.~Ahmed, Y.~Gong, S.~Hu, T.~Li and L.~Wu,
arXiv:1905.03033 [hep-th].
W.~H.~Kinney, S.~Vagnozzi and L.~Visinelli,
Class.\ Quant.\ Grav.\  {\bf 36}, no. 11, 117001 (2019)
[arXiv:1808.06424 [astro-ph.CO]].
W.~H.~Kinney,
Phys.\ Rev.\ Lett.\  {\bf 122}, no. 8, 081302 (2019)
[arXiv:1811.11698 [astro-ph.CO]].
Y.~Akrami, R.~Kallosh, A.~Linde and V.~Vardanyan,
Fortsch.\ Phys.\  {\bf 67}, no. 1-2, 1800075 (2019)
[arXiv:1808.09440 [hep-th]].
D.~Andriot,
Phys.\ Lett.\ B {\bf 785}, 570 (2018)
[arXiv:1806.10999 [hep-th]].  
S.~K.~Garg and C.~Krishnan,
arXiv:1807.05193 [hep-th]. 
A.~Kehagias and A.~Riotto,
Fortsch.\ Phys.\  {\bf 66}, no. 10, 1800052 (2018)
[arXiv:1807.05445 [hep-th]].  
F.~Denef, A.~Hebecker and T.~Wrase,
Phys.\ Rev.\ D {\bf 98}, no. 8, 086004 (2018)
[arXiv:1807.06581 [hep-th]].
K.~Choi, D.~Chway and C.~S.~Shin,
JHEP {\bf 1811}, 142 (2018)
[arXiv:1809.01475 [hep-th]].
U.~Danielsson,
JHEP {\bf 1904}, 095 (2019)
[arXiv:1809.04512 [hep-th]]. 
M.~Motaharfar, V.~Kamali and R.~O.~Ramos,
Phys.\ Rev.\ D {\bf 99}, no. 6, 063513 (2019)
[arXiv:1810.02816 [astro-ph.CO]].
H.~Fukuda, R.~Saito, S.~Shirai and M.~Yamazaki,
Phys.\ Rev.\ D {\bf 99}, no. 8, 083520 (2019)
[arXiv:1810.06532 [hep-th]].
A.~Hebecker and T.~Wrase,
Fortsch.\ Phys.\  {\bf 67}, no. 1-2, 1800097 (2019)
[arXiv:1810.08182 [hep-th]].
S.~B.~Giddings and M.~S.~Sloth,
JCAP {\bf 1101}, 023 (2011)
[arXiv:1005.1056 [hep-th]].
Schmitz:2019uti

\bibitem{Montero:2018fns} 
M.~Montero,
JHEP {\bf 1903}, 157 (2019)
[arXiv:1812.03978 [hep-th]].

\bibitem{Dvali:2018jhn} 
  G.~Dvali, C.~Gomez and S.~Zell,
  Fortsch.\ Phys.\  {\bf 67}, no. 1-2, 1800094 (2019)
  [arXiv:1810.11002 [hep-th]].


\bibitem{Bedroya:2019snp} 
A.~Bedroya and C.~Vafa,
arXiv:1909.11063 [hep-th].
\bibitem{Bedroya:2019tba} 
A.~Bedroya, R.~Brandenberger, M.~Loverde and C.~Vafa,
arXiv:1909.11106 [hep-th].



\bibitem{Dvali:2017eba} 
  G.~Dvali, C.~Gomez and S.~Zell,
  JCAP {\bf 1706}, 028 (2017)
  [arXiv:1701.08776 [hep-th]].

\bibitem{Rehman:2018nsn} 
  M.~U.~Rehman, Q.~Shafi and U.~Zubair,
  Phys.\ Rev.\ D {\bf 97}, no. 12, 123522 (2018)
  [arXiv:1804.02493 [hep-ph]].




\bibitem{Rehman:2009yj} 
  M.~U.~Rehman, Q.~Shafi and J.~R.~Wickman,
  Phys.\ Lett.\ B {\bf 688}, 75 (2010)
  [arXiv:0912.4737 [hep-ph]].



\bibitem{ArkaniHamed:2004fb} 
  N.~Arkani-Hamed and S.~Dimopoulos,
  JHEP {\bf 0506}, 073 (2005)
  [hep-th/0405159].


\bibitem{Ahmed:2019xon} 
  W.~Ahmed, A.~Mansha, T.~Li, S.~Raza, J.~Roy and F.~Z.~Xu,
  arXiv:1901.05278 [hep-ph].


\bibitem{SimonsObservatory:2018koc}
P.~Ade \textit{et al.} [Simons Observatory],
JCAP \textbf{02}, 056 (2019)
[arXiv:1808.07445 [astro-ph.CO]].


 \bibitem{Andre:2013afa}
 P.~Andre \textit{et al.} [PRISM],
 [arXiv:1306.2259 [astro-ph.CO]].

\bibitem{CMB-S4:2020lpa}
K.~Abazajian \textit{et al.} [CMB-S4],
Astrophys. J. \textbf{926}, no.1, 54 (2022)
[arXiv:2008.12619 [astro-ph.CO]].

\bibitem{Sehgal:2019ewc}
N.~Sehgal, S.~Aiola, Y.~Akrami, K.~Basu, M.~Boylan-Kolchin, S.~Bryan, S.~Clesse, F.~Y.~Cyr-Racine, L.~Di Mascolo and S.~Dicker, \textit{et al.}
[arXiv:1906.10134 [astro-ph.CO]].


  
  \bibitem{Finelli:2016cyd}
  F.~Finelli \textit{et al.} [CORE],
  JCAP \textbf{04}, 016 (2018)
  [arXiv:1612.08270 [astro-ph.CO]].

 \bibitem{Kogut:2011xw}
 A.~Kogut, D.~J.~Fixsen, D.~T.~Chuss, J.~Dotson, E.~Dwek, M.~Halpern, G.~F.~Hinshaw, S.~M.~Meyer, S.~H.~Moseley, M.~D.~Seiffert, D.~N.~Spergel and E.~J.~Wollack,
 JCAP \textbf{07}, 025 (2011)
 [arXiv:1105.2044 [astro-ph.CO]].


 





  






\bibitem{Blumenhagen:2017cxt}
R.~Blumenhagen, I.~Valenzuela and F.~Wolf,
JHEP \textbf{07} (2017), 145
doi:10.1007/JHEP07(2017)145
[arXiv:1703.05776 [hep-th]].




\bibitem{Chiang:2018lqx}
C.~I.~Chiang, J.~M.~Leedom and H.~Murayama,
Phys. Rev. D \textbf{100} (2019) no.4, 043505
doi:10.1103/PhysRevD.100.043505
[arXiv:1811.01987 [hep-th]].

f
\bibitem{Scalisi:2018eaz}
M.~Scalisi and I.~Valenzuela,
JHEP \textbf{08} (2019), 160
doi:10.1007/JHEP08(2019)160
[arXiv:1812.07558 [hep-th]].


\bibitem{Haque:2019prw}
M.~R.~Haque and D.~Maity,
Phys. Rev. D \textbf{99} (2019) no.10, 10Congradulations3534
doi:10.1103/PhysRevD.99.103534
[arXiv:1902.09491 [hep-th]].
  



\bibitem{Berera:2019zdd}
A.~Berera and J.~R.~Calder\'on,
Phys. Rev. D \textbf{100} (2019) no.12, 123530
doi:10.1103/PhysRevD.100.123530
[arXiv:1910.10516 [hep-ph]].




\bibitem{Li:2019ipk}
H.~H.~Li, G.~Ye, Y.~Cai and Y.~S.~Piao,
Phys. Rev. D \textbf{101} (2020) no.6, 063527
doi:10.1103/PhysRevD.101.063527
[arXiv:1911.06148 [gr-qc]].


\bibitem{NooriGashti:2021nox}
S.~Noori Gashti,
JHAP \textbf{2} (2022) no.1, 13-24
doi:10.22128/jhap.2021.452.1002
[arXiv:2111.06421 [gr-qc]].

 
\bibitem{Gashti:2022pvu}
S.~N.~Gashti and J.~Sadeghi,
Eur. Phys. J. Plus \textbf{137} (2022) no.6, 731
doi:10.1140/epjp/s13360-022-02961-8
  
\bibitem{Osses:2021snt}
C.~Osses, N.~Videla and G.~Panotopoulos,
Eur. Phys. J. C \textbf{81} (2021) no.6, 485
doi:10.1140/epjc/s10052-021-09283-6
[arXiv:2101.08882 [hep-th]].


\bibitem{Kamali:2021ugx}
V.~Kamali, H.~Moshafi and S.~Ebrahimi,
[arXiv:2111.11436 [gr-qc]].

\bibitem{Sabir:2019wel}
M.~Sabir, W.~Ahmed, Y.~Gong and Y.~Lu,
Eur. Phys. J. C \textbf{80} (2020) no.1, 15
doi:10.1140/epjc/s10052-019-7589-3
[arXiv:1903.08435 [gr-qc]].
  
\end{thebibliography}
\end{document}